\documentclass[aps,pra,reprint,superscriptaddress,amsmath,amssymb,nofootinbib]{revtex4-2}

\usepackage{graphicx}
\usepackage{bm}
\usepackage{hyperref}
\usepackage{physics}
\usepackage{tcolorbox}
\usepackage[T1]{fontenc}
\usepackage{amssymb}
\usepackage{amsthm}
\usepackage{physics}
\usepackage{tikz}
\usetikzlibrary{decorations.markings}
\usepackage{tcolorbox}
\usepackage{amsmath}
\usepackage{makecell}
\usepackage{array}
\newtheorem{definition}{Definition}
\newtheorem{lemma}{Lemma}
\newtheorem{corollary}{Corollary}
\newtheorem{theorem}{Theorem}
\newtheorem{observation}{Observation}
\usepackage{mathrsfs}
\usepackage{tikz-cd}

\usepackage{physics}
\usepackage{amsmath}
\usepackage{tikz}
\usepackage{mathdots}
\usepackage{yhmath}
\usepackage{cancel}
\usepackage{color}
\usepackage{siunitx}
\usepackage{array}
\usepackage{multirow}
\usepackage{amssymb}
\usepackage{gensymb}
\usepackage{extarrows}
\usepackage{booktabs}
\usetikzlibrary{fadings}
\usetikzlibrary{patterns}
\usetikzlibrary{shadows.blur}
\usetikzlibrary{shapes}

\begin{document}

\title{Tight Analysis of a One-Shot Quantum Secret
Sharing Scheme}

\author{Santanu Majhi}
\thanks{santanum\_r@isical.ac.in}
\affiliation{Indian Statistical Institute, Kolkata}

\author{Debajyoti Bera}
\thanks{dbera@iiitd.ac.in
}
\affiliation{Indraprastha Institute of Information Technology Delhi}

\date{\today}

\begin{abstract}
Quantum communication protocols can be designed to detect eavesdropping attacks, something that classical technologies are unable to do since classical information can be replicated in a non-destructive manner. 
Eavesdropping detection is, therefore, a standard feature in all the proposed quantum secret sharing (QSS) protocols. However, detection is often done by a statistical analysis of the outcome of multiple decoy rounds, and this causes a significant communication overhead.

In our quest for a QSS protocol that works even in one round, we came across a one-shot secret-sharing framework proposed by Hsu (Phys. Rev. A 2003).  The scheme was designed to 
work over public channels without requiring multiple rounds to detect eavesdropping but it lacked a thorough security analysis. In this work we present a complete characterisation of the correctness and security properties of this framework.  Our characterisation allowed us to improve the original protocol to be more resistant towards
eavesdropping. However, we prove a couple of impossibility results, including one that dictates  that complete security against an eavesdropper is not possible in this framework. Thus, it is not possible to design a perfect QSS using this framework.
\end{abstract}


\maketitle

\section{Introduction}
Secret-sharing is one of the well-known and well-studied cryptographic tasks. In a very general sense, it involves a dealer who wants to entrust multiple parties, say $N$ of them, with a common secret $S$ but does not want to put all his eggs into one basket. The solution followed in secret sharing schemes is to ``split'' $S$ into several parts ({\it aka.} shares) and share each part with a different party. The sharing should be done in a way such that any single individual, or even a set of individuals not pre-authorised by the dealer cannot gain any significant information about the secret; only an authorised group of parties will be able to reconstruct $S$ by running some algorithm on their shares.

A common manner of defining authorised and unauthorised sets use the number of parties; e.g., in $(k,N)$-threshold schemes, any group containing $k$ parties is authorised but any group with fewer parties is unauthorised. A $(2,2)$-scheme involves 2 parties and the shares of both are necessary to reconstruct $S$.

It is desirable for such schemes to be robust against eavesdropping attacks. The eavesdropper, whom we shall refer to as Eve, can be one of the parties who has gone rogue or a third-party.
It is difficult for classical schemes to prevent Eve from learning the secret, specifically, if Eve can listen to every communication between the dealer and the parties. It is also unclear how to {\em detect} if an eavesdropping has happened due to the non-intrusive nature of classical observation.
It is therefore common for classical secret sharing schemes to assume a secure channel for their communication.

However, a quantum communication channel has a few completely non-classical features that may allow protection against the above kind of vulnerabilities. First, quantum states destructively collapse upon measurement; thus, it is not possible for Eve to passively ``read'' the information sent by a dealer. Furthermore, Eve may not be able to retain a copy of a quantum state thanks to the no-cloning theorem; if Eve simply retains the transmitted qubits, the honest parties will raise a hue and cry! One of the most famous applications of this is the BB84 quantum key distribution scheme~\cite{bb84} that can detect the presence of an eavesdropper who merely tries to ``read'' information en passe.

Quantum secret sharing (QSS) extends the concept of classical secret sharing (SS) where either the secret is quantum or the protocol is quantum~\cite{CHATTOPADHYAY2024100608}. These schemes usually rely on features such as entanglement, superposition, interference, and no-cloning theorem to gain an extra mileage. 
The expectations are naturally higher for these. However, we are only aware of one theoretical model, proposed by Imai et al.~\cite{imai2003quantum}, to analyse quantum secret sharing protocols. Their model focuses on the correctness of a quantum secret sharing protocol in which all parties are ``curious-but-honest''; but it remains unclear whether that model holds in the presence of an active eavesdropper ({\it aka.} active participant) --- someone who has the ability to capture the qubits en route and even replace them with her own qubits. 

(Active) eavesdropper attacks are the worst-kinds in quantum schemes since the participants have access to additional information. If successful, such a participant can generate the secret with the involvement of fewer members than mandated; e.g., a participant who can intercept {\em all} messages between the dealer and the parties can always determine the secret on her own. A trivial approach for the honest party is to verify a reconstructed secret with the dealer. However, that either requires a private secure channel or multiple rounds (over a public channel) solely for detecting attacks. In the absence of these facilities, it may be possible to have an attack scenario in which the eavesdropper knows the correct secret, but none (the dealer and the honest parties) have spotted any anomaly. Thus, it is common for many quantum secret sharing protocols to include additional steps to explicitly prohibit participant attacks and other kinds of attacks~\cite{CHATTOPADHYAY2024100608,Kuo2023-vb}.

The common safeguards involve sending decoy states. Decoy states are common in many quantum protocols; these protocols run in multiple rounds some of which are used for checking the presence of an eavesdropper and the rest are used for the desired task~\cite{Lo_2005}. They naturally increase the cost of executing a protocol.
For example, in one of the early and one of the most-cited HBB-QSS scheme~\cite{hillery-scheme}, the dealer had to first share a GHZ state $\tfrac{1}{\sqrt{2}}[\ket{000} + \ket{111}]$ among two other parties, say Bob and Charlie, and then all three would independently measure their share in one of two non-commuting basis. Then they publicly would disclose their basis without revealing their measurement outcomes. Presence of an eavesdropper can be detected due to the disturbance an eavesdropping may cause to the outcomes. Thus, in the recommended implementation, a certain fraction of the rounds are used solely for detecting eavesdropping.



The current solutions beg an important question: {\em Is it possible to design a one-shot QSS scheme?} One that detects eavesdropping even in one round.

The motivation behind this work is an early QSS scheme proposed by Hsu~\cite{hsu-scheme}, which we refer to as H03-QSS, that was designed to detect eavesdropping without using decoy rounds~\footnote{In the words of its author, ``Detection is not based on the statistical violation of an outcome sequence, but on the correlation of the outcomes of a qubit pair.".}. Interestingly, the scheme could be used a framework, i.e., it allows variations in the form of choosing different sets of ``nonce'' states~\footnote{We use nonce to mean states chosen randomly but not kept secret, i.e., eventually disclosed. Similar to their role in secure protocols, they improve security by adding randomness.} Given that most of the proposed QSS schemes use multiple rounds to detect eavesdropping, the one-shot nature of H03-QSS appears to be an attractive blueprint for QSS schemes.

However, Hsu proposed the scheme without an adequate proof of security. Our primary objective was to tightly characterise the security of the H03-QSS scheme. Interestingly, we made a few surprising discoveries en route, including some negative results that fundamentally limit the security achievable within this framework.

\begin{table}[h!]
\centering
\renewcommand{\arraystretch}{1.1}
\setlength{\tabcolsep}{4pt}
\begin{tabular}{lcc}
\toprule
 & \textbf{Original nonces} & \textbf{Our nonces} \\
\midrule
Recoverable                   & Yes       & Yes       \\
Secrecy attack                & Resistant & Partial   \\
Intercept-measure-resend      & Resistant & Resistant \\
Intercept-fake-resend         & Not       & Partial   \\
\bottomrule
\end{tabular}
\caption{Comparison of the original nonces {\it vs.} our nonces with respect to their resistance towards different attacks.} 
    \label{tab:comparison_nonces}
\end{table}



\subsection{Overview of Results}


We will give a brief overview of H03-QSS here and explain it in detail in a later section. H03-QSS is executed in multiple stages. Apart from the secret $s$, the dealer commits to a random chosen nonce state, say $c$, in the first stage. Then the dealer constructs a state $\ket{\psi_{c,s}}$ using the secret $s$ and nonce $c$, and shares it with the parties. $\ket{\psi_{c,s}}$ is constructed such that it reveals nothing about $s$. Later, the dealer discloses $c$, allowing authorized parties to recover $s$. The immediate observation is that no encoding of $s$ is ever transmitted, reducing the risk of eavesdropping. A natural question, thus, arises: {\em Is it possible to prove that H03-QSS is indeed secure against an active rogue participant?}

The nonces recommended by Hsu are inspired by the reflection operators employed by the Grover's search algorithm (see Table~\ref{table:states}). Our observation is that the role of the nonce states is central to the correctness and security guarantee of the scheme. Thus, even if the proposed nonces turn out to be vulnerable~\cite{hao2010eavesdropping}, could one choose a different set of nonces that retains both the one-shot nature of the original protocol and the security claimed about it?

Our overall discovery is that a one-shot QSS over public channels using Grover-type reflections is impossible without compromising on certain aspects of security. Following are the specific technical results that we show.

\begin{enumerate}
    \item (Section~\ref{sec:characterization}) We tightly characterise the correctness and security requirements of H03-QSS in terms of various properties of the nonces. Our characterisation is information theoretic in nature and does not rely on complexity theoretic assumptions. The following results follow from our characterisation.
    
    \item We demonstrate that the original set of nonces in H03-QSS are vulnerable to eavesdropping (Section~\ref{sec:analysis_original_Hsu}); this generalizes an attack on H03-QSS demonstrated by Hao et al.~\cite{hao2010eavesdropping}. In fact, we show how to generate an attack for any set of weak nonces.
    
    \item We prove two no-go results. The first one is that 
    Eve always has a $\tfrac{1}{4}$ probability of remaining undetected after an intercept-fake-resend attack no matter what nonces are used~(Section~\ref{subsec:no-attack-not-possible}).
    
    \item The second one is worse. If the scheme is designed to ensure secrecy, i.e., that the secret cannot be guessed by any individual party honestly running the protocol, then Eve can always launch an (intercept-fake-resend) attack to successfully learn the secret (Section~\ref{subsec:secrecy-implies-imr}). Thus, to increase the chance of detecting this attack (which itself is bounded above by $3/4$), it is necessary to allow the honest parties a small chance to guess the secret. 
    
    \item The negative results imply that secrecy must be compromised to prevent other attacks. But to what extent? We present a set of nonces and prove that both secrecy and protection against intercept-fake-resend attack can be partially attained~(Section~\ref{section:analysis_our_scheme}). Table~\ref{tab:comparison_nonces} presents a comparison of our nonces with the original ones.
    

    \item The original scheme was developed only for classical 2-bit secrets, or 1-bit secret along with cheat-detection. On a positive note, we are able to show that H03-QSS can also be used to send arbitrary quantum states as secret (Observation~\ref{obs:1}).

\end{enumerate}

The immediate advantage of our characterization is to do away with case-based analysis and hit-and-trial methods to demonstrate the correctness and security of H03-QSS (see, for example, the exhaustive tables in the related works~\cite{hsu-scheme,hao2010eavesdropping}). 
For example, we show that it is necessary and sufficient for the amplitudes of the basis states in any nonce to be \textonehalf~for a secret to be recoverable by the honest parties, thus eliminating the need for exhaustive case analysis across all secrets and nonces.


\subsection{Related Work}

Hillery et al.~\cite{hillery-scheme} introduced the initial quantum secret sharing (HBB-QSS) scheme in 1999, utilizing the Greenberger-Horne-Zeilinger (GHZ) state. In that scheme, the dealer used ideas similar to the BB84-QKD protocol to create a shared secret quantum state with 2 other parties; however, it did not allow the dealer to send a secret of his choice. 

The scheme that we will analyse and improve upon is a (2,2)-scheme that allows a dealer to send a classical 1-bit secret of his choice; the earliest (2,2) schemes to share a desired secret were designed by Cleve et al.~\cite{cleve-qss}.



Proving security of multiparty computation is in general difficult due to multiple points of vulnerability, and quantum schemes always bring in additional abilities of adversarial parties. Unfortunately, most quantum ``secure'' protocols are often proposed without a formal proof of security leading to later discoveries of vulnerabilities.

Take for instance the HBB-QSS scheme mentioned earlier~\cite{hillery-scheme}. Soon after the scheme was proposed, Karlsson et al.\ explained how a dishonest player may evade detection by attacking the order in which the messages in that scheme were announced~\cite{Karlsson-PhysRevA.59.162}. Unfortunately, most QSS schemes that have been proposed are based on the framework used in HBB-QSS, and many of them, like HBB-QSS, are vulnerable to participant attacks~\cite{Karlsson-PhysRevA.59.162,Bai2021-wn,HBB-Attack}.

Another framework for QSS has been entanglement-swapping followed by measurements in a carefully-chosen basis~\cite{Zhang_Ying-Qiao_2006}. However, such schemes are also known to suffer from participant attacks~\cite{Song_Ting-Ting_2009}.



The H03-QSS scheme proposed by Hsu~\cite{hsu-scheme}, that we analyse in this work, is also known to be vulnerable to participant attack. This was later pointed out by Hao et al.~\cite{hao2010eavesdropping} whose attack crucially used the properties of the nonces proposed by Hsu.
To fix the scheme they proposed that decoy states can be used. Our intention in this work is to explore the possibility of designing better nonces such that eavesdropping can be detected as originally envisioned, without decoy states.


Several works have explored extensions and improvements of Hsu's protocol.
Tseng et al.~\cite{Tseng2012QuantumSecretSharing,Yu_2022} observed that H03-QSS, like many quantum schemes for secret sharing, requires the parties to store quantum states for a long time and apply quantum operations on their share. Thus, they proposed a variation that combined the HBB-QSS style of generating a classical secret by measuring states in superpositions and involving the Grover reflection operators employed by Hsu. Similar to HBB-QSS, they also use decoy states to detect eavesdropping.

There has been several interesting directions that research on QSS is leading into. Several research groups are investing into efficient hardware implementations of QSS schemes~\cite{PhysRevResearch.6.033036} while some others are studying the performance in a noisy environment~\cite{Basak2023-wb}. However, Our work is of a theoretical nature and

If H03-QSS was to be implemented and deployed on the currently available hardware, it would be important to analyse its correctness and security guarantees in a NISQ environment, similar to how Basak et al.~\cite{Basak2023-wb} analyse the HBB-QSS and two other QSS schemes. We left out such error analysis in this work, partly due to the theoretical limitations of H03-QSS that we discovered.


\section{Background}


\begin{table}[t]
\begin{ruledtabular}
\begin{tabular}{c|c|c|c}
$\ket{+}$ & $\ket{-}$ & $\ket{+i}$ & $\ket{-i}$ \\
\hline
$\tfrac{1}{\sqrt{2}}\!\left( \ket{0} + \ket{1} \right)$ &
$\tfrac{1}{\sqrt{2}}\!\left( \ket{0} - \ket{1} \right)$ &
$\tfrac{1}{\sqrt{2}}\!\left( \ket{0} + i\ket{1} \right)$ &
$\tfrac{1}{\sqrt{2}}\!\left( \ket{0} - i\ket{1} \right)$ \\
\end{tabular}
\end{ruledtabular}
\caption{\label{table:states}States used to construct nonces.}
\end{table}

\begin{table}[t]
\centering
\renewcommand{\arraystretch}{1.3}
\setlength{\tabcolsep}{4pt}
\begin{tabular}{cc|cc}
\toprule
$\ket{+}\ket{+}$   & $\ket{+}\ket{-}$   & $\ket{+}\ket{+i}$   & $\ket{+}\ket{-i}$   \\
$\ket{-}\ket{+}$   & $\ket{-}\ket{-}$   & $\ket{-}\ket{+i}$   & $\ket{-}\ket{-i}$   \\
\bottomrule
$\ket{+i}\ket{+}$  & $\ket{+i}\ket{-}$  & $\ket{+i}\ket{+i}$  & $\ket{+i}\ket{-i}$  \\
$\ket{-i}\ket{+}$  & $\ket{-i}\ket{-}$  & $\ket{-i}\ket{+i}$  & $\ket{-i}\ket{-i}$  \\
\bottomrule
\end{tabular}
\caption{Nonces used in the original H03-QSS protocol.}
\label{tab:hsu_nonces}
\end{table}


\subsection{Grover-search Operator}
\label{subsec:grover-operator}
The design of H03-QSS is based on an interesting observation made by its author about the ``reflection'' operators used in the Grover's algorithm.

 Suppose $\ket{\alpha}$ is some 2-qubit state. The reflection operator is defined below.
$$U_{\ket{\alpha}}=I-2\ket{\alpha}\bra{\alpha}$$
Observe that $U_{\ket{\alpha}}^\dagger = U_{\ket{\alpha}}$. If $s$ is a bit string, $U_s$ may be written instead of $U_{\ket{s}}$.

The behaviour of this operator can be understood by its action on any (orthonormal) basis that contains $\ket{\alpha}$:
$$U_{\ket{\alpha}} \ket{\beta} = 
\begin{cases}
    -\ket{\alpha} & \beta = \alpha\\
    \ket{\beta} & \braket{\beta}{\alpha} = 0
\end{cases}
$$

The Grover's algorithm for unordered search in an array of 4 elements with 1 solution essentially uses the identity:
\begin{equation}
    -U_{c} U_s \ket{c} = \ket{s},
    \label{eq: grover_reflection}
\end{equation}
where $\ket{c} = \tfrac{1}{2}[\ket{0} + \ket{1}]^{\otimes 2} = \ket{++}$, and $s$ is a any 2-bit string. The author observed that the identity extends to other initial states (up to some global phase) chosen from the set $\mathcal{I}=\{\ket{x} \otimes \ket{y} ~:~ x,y \in \{+,-,+i,-i\} \}$ (refer to Table~\ref{table:states} for description of the states).

This operator is an integral part of the H03-QSS scheme that we describe in the next section.

 \setlength\intextsep{1pt}
\bigskip
\begin{table}[!t]
\renewcommand{\arraystretch}{1.17}
\begin{ruledtabular}
\begin{tabular}{l|l || l|l}
\textbf{Notn.} & \textbf{Description} & \textbf{Notn.} & {Description} \\
\hline
$m$ & mode & ${J}$ & {proposed nonce set} \\
$\ket{\alpha}$ & fake state shared by Eve & $\mathfrak{Sec}$ & secret bit to be sent \\
$\ket{\psi_i}$ & nonces in ${J}$ & $\rho$ & $\ket{\alpha}\bra{\alpha}$ \\
$s$ & 2-bit secret & $U_{x}$ & $I - 2\ket{x}\bra{x}$ \\
$\rho^B$ & $Tr_{E}(\rho)$ & $\mathcal{I}$ & original nonce set \\
$\ket{\psi_{i,s}}$ & $U_s \ket{\psi_i}$ (sent by dealer) & $\sigma_{i,s}$ & $\ket{\psi_{i,s}}\bra{\psi_{i,s}}$ \\
$b$ & regenerated secret & $\sigma_{i,s}^B$ & $Tr_{E}(\sigma_{i,s})$ \\
$V_i$ & Eve's local unitary & \\
\end{tabular}
\end{ruledtabular}
\caption{\label{tab:notation}List of notations used in this work.}
\end{table}

\subsection{Grover-based H03-QSS by Hsu}
\label{subsec:hsu}
\begin{figure}[htbp]
\centering
\begin{minipage}[t]{\linewidth}
    \centering
    \includegraphics[width=1.07 \linewidth]{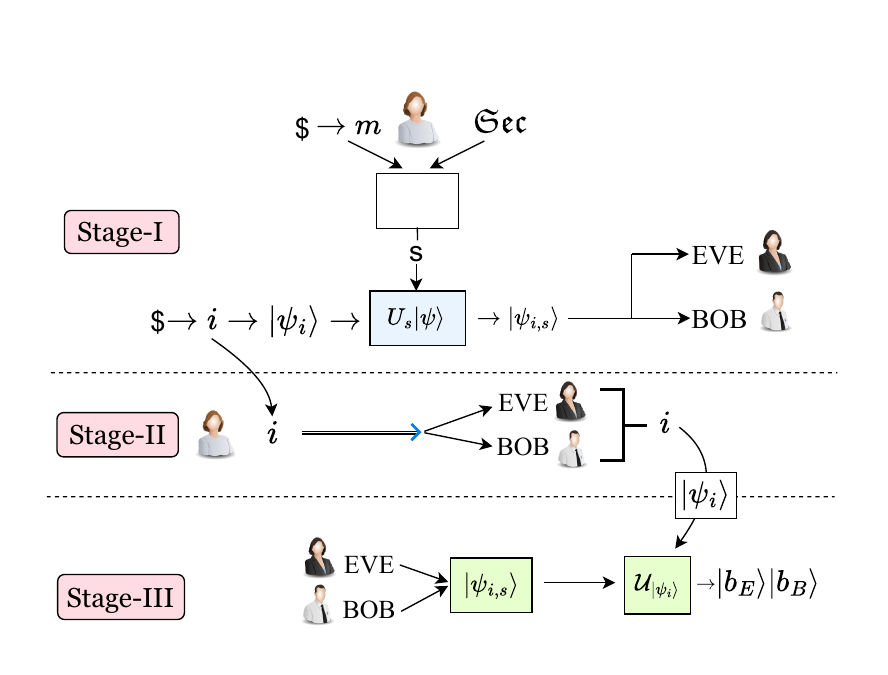}
    \caption{\label{fig:hsu}
 The first three stages of the protocol when Eve is acting honestly. The symbol \$ indicates drawing from a uniform random distribution, and $\Longrightarrow$ indicates communication over a public channel.}
\end{minipage}\hfill
\end{figure}

In this subsection, we will give a quick overview of the H03-QSS scheme~\footnote{Hsu denoted the nonce by $s$ for which we use $\ket{\psi}$, and he denoted the secret by $w$ for which we use $s$.}. Let $\mathfrak{Sec} \in \{0,1\}$ denote the secret a dealer has to share among two parties named as Eve and Bob~\footnote{As is the norm, Bob is always curious-but-honest.}.
%
A schematic of the protocol is illustrated in Figure~\ref{fig:hsu}.

\begin{enumerate}
    \item[] {\bf Stage-I:} (Distribution of Shares)
    \item Dealer choose a mode $m \in \{\mathtt{SECRET, DETECT}\}$ uniformly at random.
    \item Dealer maps $\mathfrak{S}$ and $m$ to a 2-bit string $s$ according to a known mapping $$\mathcal{M}:\{0,1\} \times \{\mathtt{SECRET, DETECT}\} \to \{0,1\}^2.$$
     In the original protocol, the following mapping was used which we will continue to use.
    \begin{itemize}
        \item $\mathcal{M}(\mathfrak{S},\mathtt{SECRET}) \mapsto \mathfrak{S}\mathfrak{\bar{S}}$.
        \item $\mathcal{M}(\mathfrak{S},\mathtt{DETECT}) \mapsto \{00,11\}$ uniformly at random.
        \item Thus, $s=\mathfrak{S}\mathfrak{\bar{S}}$ with probability \textonehalf, $00$ with probability \textonequarter, and $11$ with probability \textonequarter.
    \end{itemize}
    \item A two-qubit nonce $\ket{\psi_i}$ is chosen from a known set of nonces. In the H03-QSS proposed by Hsu, the nonces are chosen from $\mathcal{I}$ defined in Sec.~\ref{subsec:grover-operator}.
    \item Dealer generates the two-qubit state $\ket{\psi_{i,s}} = U_s\ket{\psi_i}$ and sends one qubit to Eve and another to Bob (both the qubits are symmetrical in nature).
    \item After both the parties acknowledge receiving their shares (over a public channel), the dealer moves to Stage-II.\\[0.3em]
    \item[] {\bf Stage-II:} (Nonce Announcement)
    \item Dealer announces (over a public channel) the choice of nonce $\ket{\psi_i}$.\\[0.3em]
    \item[] {\bf Stage-III:} (Secret Recovery)
    \item When the parties want to regenerate the secret, they get together and apply $U_{\ket{\psi_i}}$ to their joint-state.
    \item Then they perform a measurement in the standard-basis; suppose, the outcome is $\ket{b}$ for $b \in \{00,01,10,11\}$.\\[0.3em]
    \item[] {\bf {Stage-IV:}} (Reconciliation)
    \item A cheat-detection strategy is executed based on $s$. The strategy followed in the original protocol is described next. Let $b_E$ and $b_B$ denote the first and second bits of $b$, respectively.
    \item If $b_E \not= b_B$, i.e., $b \in \{01,10\}$, consider $b$ as the regenerated secret, and $b_E$ as $\mathfrak{Sec}$. All the (good) folks happily retire.
    \item If $b_E=b_B$, i.e., $b \in \{00,11\}$, parties publicly announce $b$ to the dealer, who follows the next steps.
    \begin{itemize}
        \item If $m=\mathtt{DETECT}$, and $s=b$, announce over classical channel to drop this round (eavesdropper presence is not detected, but $\mathfrak{Sec}$ was not shared either).
        \item If $m=\mathtt{DETECT}$, and $s\not=b$, announce over classical channel the presence of an eavesdropper.
        \item If $m=\mathtt{SECRET}$, announce over classical channel the presence of an eavesdropper.
    \end{itemize}
    \item If $m=\mathtt{DETECT}$ but the parties did not announce their generated secrets, dealer announces over classical channel the presence of an eavesdropper.
    \item The above steps are applicable to the mapping $\mathcal{M}$ defined in the original protocol; they can be easily adapted for any other mapping without impacting security guarantees in any manner.
\end{enumerate}


If both parties are honest, they will generate $b=s$ in Stage-3 since
$$U_{\ket{\psi_{i}}} \ket{\psi_{i,s}} = U_{\ket{\psi_i}} U_s \ket{\psi_i} = \ket{s}\quad\mbox{ as per Eq.~\ref{eq: grover_reflection}}.$$

$s$ can be used to derive the actual secret bit $\mathfrak{Sec}$, so we will focus only on generating $s$ and call it the ``secret''.

The cheat-detection steps detect the presence of an eavesdropper by following this rule-of-thumb.
\begin{itemize}
    \item In the detection mode, the dealer expects that the parties will announce a generated secret $b$ that exactly matches $s$.
    \item In the secret message mode, the dealer expects the parties {\em not to} announce anything at all, i.e., $b$ is neither 01 or 10.
\end{itemize}

Earlier, Imai et al.~\cite{imai2003quantum} had proposed an information-theoretic model of QSS. 
%
A correct QSS protocol has to satisfy two conditions of correctness: (1) Recoverability -- the players in any set of authorized players should be able to gain complete information about the secret; this was shown by Rietjens to be equivalent to the existence of a mapping that can regenerate $S$~\cite{rietjens2004quantum}. (2) Secrecy -- the players of any unauthorized group should {\em not} be able to extract significant information about the secret.
%
%
The H03-QSS scheme can be proved correct according to Imai's characterization. 
Unfortunately, Imai et al.'s model is not applicable in the presence of an eavesdropper.

\subsection{Attack against H03-QSS (\cite{hao2010eavesdropping})}
\label{subsec:hao}

\begin{figure}[htbp]
\centering
\begin{minipage}[t]{\linewidth}
    \centering
    \includegraphics[width=\linewidth]{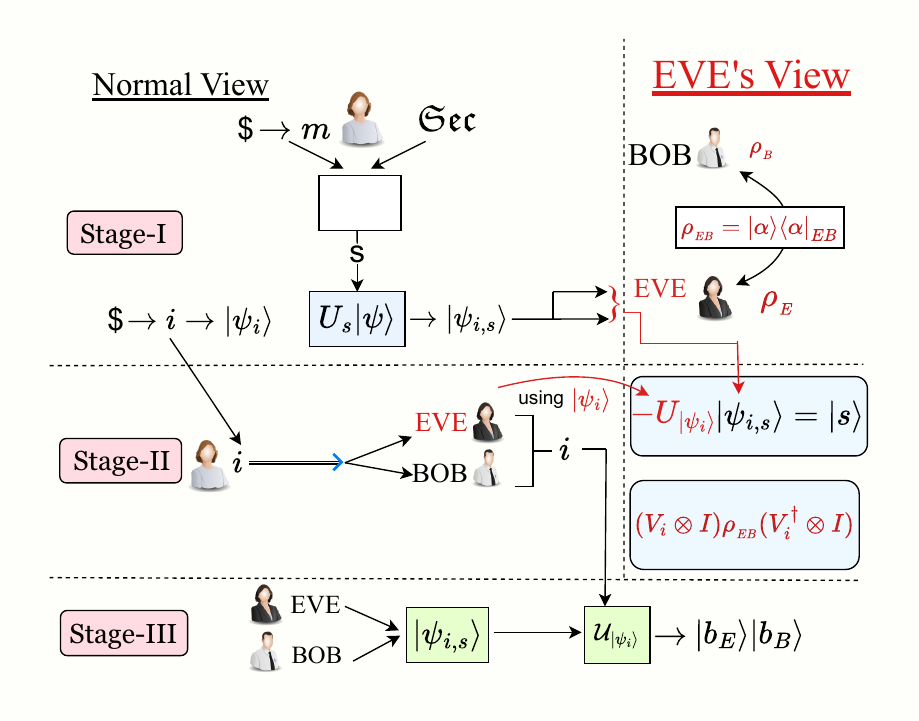}
    \caption{Eavesdropping attack against H03-QSS. Observe that Eve has successfully generated $\ket{s}$ in Stage-II itself without Bob's involvement.}
    \label{fig:hao}
\end{minipage}\hfill
\end{figure}

In this section we discuss the idea of the attack discovered by Hao et al.~\cite{hao2010eavesdropping}; a schematic of the attack is presented in Fig.~\ref{fig:hao}.

As is common in active participant attacks, for this attack we assume that Eve has the ability to intercept the message from the dealer to Bob as well as send message to Bob pretending to be the dealer.
%
To launch an attack, Eve first prepares the 2-qubit state $\ket{\alpha} = \tfrac{1}{\sqrt{2}}[\ket{01}+ \ket{10}]$.

In Stage-I, when the dealer transmits the two qubits, say in state $\ket{\psi_{i,s}}$, Eve captures both and keeps them with her. She immediately sends Bob the second qubit of $\ket{\alpha}$ pretending to be the dealer. The state $\ket{\alpha}$ does not depend on $\ket{\psi_{i,s}}$.

In Stage-II, after the dealer announces the nonce $\ket{\psi_i}$, Eve applies $U_{\ket{\psi_i}}$ on $\ket{\psi_{i,s}}$ which is the secret-reconstruction procedure, and obtains $\ket{s}$. That's it --- Eve has found the secret.

However, if Bob and Eve were to come together in Stage-III to generate the secret, the would apply $U_{\ket{\psi_i}}$ to $\ket{\alpha}$ and the resultant state may not pass the cheat-detection steps of the protocol. Thus, to avoid detection, in Stage-II Eve applies some $V_i$, chosen according to $\ket{\psi_i}$ and $s$, on her share of $\ket{\alpha}$ such that
$$(V_i \otimes I) \ket{\alpha} = (V_i \otimes I) \tfrac{1}{\sqrt{2}}[\ket{01} + \ket{10}] = U_s\ket{\psi_i} = \ket{\psi_{i,s}}.$$

The reader may refer to Hao et al.~\cite{hao2010eavesdropping} for the exhaustive list of $V_i$ for each $\ket{\psi_i}$ and $s$.

Thus, at the end of Stage-II, Eve has managed to create a joint state with Bob which is exactly the state the dealer wanted to share with both of them. So, when Bob and Eve want to reconstruct the secret, the reconstruction step works as expected and they are able to recreate the original secret. All the cheat-detection steps naturally fail to detect this tampering.

The attack leaves much to ponder upon. Is the state $\ket{\alpha}$ and the set of operators $\{V_i\}$ unique, or it is possibly to construct them from the nonces? Is it possible to launch the attach using some different $\ket{\alpha}$?

\section{Mathematical Characterization}\label{sec:characterization}

A desirable QSS scheme must satisfy three conditions: recoverability, secrecy and security.

Recoverability indicates the probability with which an authorised set (here, both the parties) can create the same secret used by the dealer. If the probability is 1, then the secret is perfectly recoverable.

In the absence of active participant attacks (Eve is only honest but curious), secrecy is defined as the protection against unauthorized parties (here Eve) knowing the secret (without Bob's help). Unfortunately, secrecy is fundamentally impossible to achieve in any active participant attack since Eve has to simply intercept both the shares, send any state to Bob, and then when the dealer announces the nonce, run the secret recovery operations on her intercepted qubits to reconstruct the dealer's secret. It is easy to see that recoverability and secrecy both cannot be satisfied at the same time. 

A scheme is said to be secure if it holds its ground against different types of attacks. Since participant attacks are strictly more powerful compared to external attacks, we will only discuss the former.

In all active participant attacks, Eve, the active participant, is allowed the ability to intercept all the shares in Stage-I, i.e., she gets hold of the state $\ket{\psi_{i,s}}$. Next she create a 2-qubit state $\ket{\alpha}$; this state could either be created after measuring $\ket{\psi_{i,s}}$ in some basis or could be chosen from a fixed family of states. Eve sends one qubit of $\ket{\alpha}$ to Bob.

Attacks involving a measurement to create $\alpha$ are known as {\em intercept-measure-resend}, while those in which Eve creates $\ket{\alpha}$ without measurement (thus, retaining $\ket{\psi_{i,s}}$ as it is) are called as {\em intercept-fake-resend}.

After Bob and Eve informs the dealer about receiving their shares, the dealer announces the nonce $\ket{\psi_i}$ in Stage-II. This automatically prevents any possibility of timing attack. 
At this point, Eve can apply some unitary, possibly $U_{\ket{\psi_i}}$, to her qubits. She does it hoping to set the joint-state of her and Bob's qubits to some state.

Next, in the secret recovery phase (Stage-III), Eve and Bob get together and both present their qubits on which they apply $U_{\ket{\psi_i}}$. Eve hopes that the outcome will pass the cheat detection steps in Stage-IV.

Given that secrecy is not possible in a perfectly recoverable scheme, we can at best require that any form of eavesdropping be detected in a secure QSS scheme.

In the following subsections we derive mathematical conditions for the correctness and security of the H03-QSS scheme but with an arbitrary set of nonces denoted $J$; the number of nonces will be denoted $k$. Earlier works by Hsu~\cite{hsu-scheme} and Hao et al.~\cite{hao2010eavesdropping} did not provide a general characterization, and that made it difficult to rigorously analyse their approaches and study their enhancements.

\subsection{Recoverability}
\label{subsec:recoverability}


For H03-QSS, the secret recovery step involves applying $U_{\ket{\psi}}$ on the shared state $U_s\ket{\psi}$ for every nonce $\ket{\psi} \in J$. This gives us the following mathematical definition of recoverability.

\begin{definition}[Recoverability] An H03-QSS scheme is said to be recoverable if for every secret $s$ and every nonce $\ket{\psi} \in J$, $s$ can be recovered by both the parties following the steps of Stage-III, i.e., 
\begin{align}
U_{\ket{\psi}} U_s\ket{\psi} &= \ket{s}~\text{(up to some phase)}, \nonumber \\
\text{or equivalently,}\quad 
U_{\ket{\psi}} \ket{s} &= U_s \ket{\psi}~\text{(up to some phase)}.
\label{eqn:recoverability}
\end{align}
\end{definition}

This unfortunately requires checking all combinations of secrets and nonces --- an approach undertaken by Hsu (see Table I of \cite{hsu-scheme}). We now present an equivalent but simpler condition that is easy to check.

\begin{lemma} 
An equivalent condition for Equation~\ref{eqn:recoverability} is that
    $$|\braket{s}{\psi}|=\tfrac{1}{2}$$
    holds for all secret $\ket{s}$ and all nonce $\ket{\psi} \in J$.
    \label{lemma:recoverability_condition}
\end{lemma}

The proof is available in Appendix~\ref{appendix:lemma_recoverability_proof}. An interesting aspect of the proof is that it holds for any 2-qubit state $\ket{\psi}$, unlike only the standard basis states ($\ket{00}, \ket{01}, \ket{10}, \ket{11}$) from which Hsu chose his secrets.

\begin{observation}
    An H03-QSS scheme remains recoverable for any 2-qubit state $\ket{s}$ satisfying Lemma~\ref{lemma:recoverability_condition}. \label{obs:1}
\end{observation}

This allows us to extend H03-QSS to any 2-qubit state as secret. Of course, the secrets need to additional meet the requirements of secrecy and security.

\subsection{Secrecy}\label{subsec:defn_secrecy}
Another correctness condition of a QSS scheme is secrecy --- how safe is a share with any curious but honest party. A secret QSS scheme should not allow such a party to know the secret without the involvement of the other party.

\begin{definition}[Secrecy]
Secrecy of a party is defined as the probability with which that party, in a curious but honest scenario, {\em cannot} guess a randomly chosen $\mathfrak{S}$ on its own.
\end{definition}

Observe that secrecy is at most \textonehalf, since any party can always {\em guess} $\mathfrak{S}$ by randomly tossing a coin, but this is the best possible value. Secrecy of 0 indicates that the secret can be correctly inferred.

The characterization below uses Eve as the honest but curious party, but an equivalent characterization exists for Bob as well. 
%
An honest but curious Eve can neither intercept Bob's qubits nor send him fake qubits. Thus, to launch an attack, Eve has to construct the secret $s$ based on her share of $\ket{\psi_{i,s}}$ and the nonce $\ket{\psi_i}$ announced in Stage-II. 

According to the mapping $\mathcal{M}$ used in the original protocol, $s$ is chosen as:
\[ 
\mathfrak{S} \mathfrak{\bar{S}} \text{ with prob. \textonehalf},~ 00~ \&~  11 \text{ with prob. \textonequarter\ each}.
\]

Thus, the secret state $\ket{\psi_{i,s}}$ is chosen according to:
\[
\ket{\psi_{i,\mathfrak{S} \mathfrak{\bar{S}}}} \text{ with prob. \textonehalf},~ \ket{\psi_{i,00}} \&  \ket{\psi_{i,11}} \text{ with prob. \textonequarter\ each}.
\]

Therefore, after the nonce $\ket{\psi_i}$ is announced, Eve's share, denoted $\lambda E^{\mathfrak{S}}_i$, is the partial trace of the above mixture. 
We use the notation $\sigma_{i,s}$ to denote $\ket{\psi_{i,s}}\bra{\psi_{i,s}}$.

\begin{align*}
    \lambda E^{\mathfrak{S}}_i & = \Tr_B{\lambda^{\mathfrak{S}}_i}, \quad \text{where,}\\
    \lambda^{\mathfrak{S}}_i  &= \tfrac{1}{2} \ketbra{\psi_{i,\mathfrak{S} \mathfrak{\bar{S}}}} + \tfrac{1}{4} \ketbra{\psi_{i,00}} + \tfrac{1}{4} \ketbra{\psi_{i,11}}\\
    &= \tfrac{1}{2} \sigma_{i,\mathfrak{S} \mathfrak{\bar{S}}} + \tfrac{1}{4} \sigma_{i,00} + \tfrac{1}{4} \sigma_{i,11}.
\end{align*}

To guess $\mathfrak{S}$, Eve can distinguish between $\lambda E^0_i$ and $\lambda E^1_i$, and the probability of this is upper bounded by $$\tfrac{1}{2} + \tfrac{1}{4} \| \lambda E^0_i - \lambda E^1_i \|_1 = \tfrac{1}{2} + \tfrac{1}{8} \| \Tr_B (\sigma_{i,01} - \sigma_{i,10}) \|_1.$$

\begin{lemma}\label{lemma:eve-guessing-secret}
    The probability of Eve's ability to guess $\mathfrak{S}$ for any nonce $\ket{\psi_i}$ may be calculated as
    $$ \tfrac{1}{2} + \tfrac{1}{8} \| \sigma^B_{i,01} - \sigma^B_{i,10} \|_1. $$
\end{lemma}

Thus, if $\sigma^B_{i,01} = \sigma^B_{i,10}$, then Eve cannot meaningfully guess $\mathfrak{S}$. The lemma allows us to characterise secrecy as the expected value of one minus the above quantity (since we are interested in the event of not being able to guess) where the expectation is taken over the choice of the nonce.

\begin{lemma}\label{lemma:secrecy}
    The secrecy with respect to Eve may be calculated as $\tfrac{1}{2} - \tfrac{1}{8|J|} \sum_i \| \sigma^B_{i,01} - \sigma^B_{i,10} \|_1$.
\end{lemma}




\subsection{Intercept-Measure-Resend Attack}
In this attack, Eve intercepts the qubit meant for Bob, giving her both qubits of $\ket{\psi_{i,s}}$, and these she then decides to measure in an attempt to get some information about $s$. Further, to avoid detection, she must ensure that the protocol plays according to the script, i.e., (i) she must send some qubit to Bob, say in {state} $\rho_B = \Tr_E(\rho_{EAB})$ pretending it to come from the dealer, (ii) she must present a qubit in a suitable state, say in state $\rho_E = \Tr_{AB}(\rho_{EAB})$, for the secret recovery step jointly executed by her and Bob. Here,  $\rho_{EAB}$ denotes the joint-state of a register that Eve prepares for this purpose; the register may include additional qubits that we indicate using the subsystem $A$, apart from one (subsystem $B$) to send to Bob and another (subsystem $E$) to send during the secret recovery phase.

Hsu, in his paper, considered a common choice for the measurement basis based on a guess $\ket{\psi_j}$ of the nonce. Eve would then apply $U_{\ket{\psi_j}}$ to attempt recovery, yielding some secret $s'$ (possibly incorrect). To avoid detection, she prepares $\ket{\psi_{j,s}} = U_{s'} \ket{\psi_j}$, keeps one qubit, and sends the other to Bob. The leftover qubit of $\ket{\psi_{j,s}}$ would be used for the secret recovery stage. %
If her guess $j$ equals $i$, then, of course, the joint-state of Eve and Bob would be $\ket{\psi_{i,s}}$, and not only Eve would know the secret, no one will have any clue of her attack. Thus, the probability of success of this attack is the probability that Eve is able to generate the original secret $s$ better than a random guessing.
%
%
Hsu demonstrated that his scheme was secure by displaying an exhaustive case analysis for the original set of nonces ($\mathcal{I}$). 
However, we can characterize the 
probability that Eve will observe $s$ in a much simpler manner 
using Eq.~\ref{eqn:recoverability}:
\begin{align*}
\left| \braket{s}{U_{\ket{\psi_j}} \ket{\psi_{i,s}}} \right|^2 
&= \left| \braket{s}{U_{\ket{\psi_j}} U_s \ket{\psi_i}} \right|^2 \nonumber \\
&= \left| \braket{\psi_j}{U_{s} U_s \ket{\psi_i}} \right|^2 \nonumber \\
&= \left| \braket{\psi_j}{\psi_i} \right|^2 .
\label{eq:overlap}
\end{align*}
This allows us to compute the success probability of guessing using only the nonces in a more elegant manner.

Interestingly, the above analysis is applicable only when Eve follows the prescription described by Hsu, i.e., Eve first guessing the nonce. Other possibilities are left open? We observe if the state of the qubits intercepted by Eve is independent of the secret bit ${\mathfrak{Sec}}$, then Eve would not be able to extract any meaningful information about ${\mathfrak{Sec}}$ through a measurement. This allows us to derive a sufficient condition for protection against any form of intercept-measure-resend attack.

\begin{lemma}\label{lemma:imr}
    An H03-QSS scheme ensures protection against intercept-measure-resend attacks if 
    \begin{align*}
    \tfrac{1}{|J|}\sum_{i} \lambda_i^0 &= \tfrac{1}{|J|}\sum_{i} \lambda_i^1,~\text{or equivalently,}\\
    \sum_i \sigma_{i,01} &= \sum_i \sigma_{i,10}.
    \end{align*}
\end{lemma}

$\lambda^\mathfrak{S}_i$ was defined in Section~\ref{subsec:defn_secrecy} as the state of the shares distributed by the dealer when secret bit is $\mathfrak{S}$ and nonce is $\ket{\psi_i}$.



\subsection{Intercept-fake-resend Attack}
\label{sec:internal-tamperability}

We discussed this attack in Section~\ref{subsec:hao} earlier. This is the strongest of the attacks that we consider in this work in the sense that Eve is always able to determine the secret. Thus, the best protection is to ensure that Eve's attack is detected in the Reconciliation Stage.




On the other hand, from Eve's perspective, if she is able to transform $\ket{\alpha}$ to exactly $\ket{\psi_{i,s}}$ then her tampering cannot be detected any further. Furthermore, even if $\ket{\alpha}$ is transformed to a state close to $\ket{\psi_{i,s}}$, the key recovery step will generate the original secret with a reasonable probability. Much to Eve's delight, the Reconciliation Stage will successfully pass.

Thus, we will be concerned only about ``Eve's tampering leads to generation of $s$ in Stage-III'', an event that we will denote $\mathcal{E}(s)$ and that indicates a weakness of the scheme.




{The next lemma} derives the probability of $\mathcal{E}(s)$.

\begin{lemma}
  {Fix any secret $s$.
    Let $\sigma_{s,i}$ denote the density operator $U_s \ket{\psi_i}\bra{\psi_i}U_s$ where $s$ is some secret and $\ket{\psi_i}$ is some nonce. Let $\rho$ denote any two-qubit state $\ket{\alpha}\bra{\alpha}$. Let $\{V_1, V_2, \ldots V_k\}$ denote the single-qubit unitary operators Eve applies corresponding the different nonces ($V_i$ may depend on $\sigma_{s,i}$ and $s$)}. 

    \begin{equation}\mbox{Then, }
    \Pr_{\ket{\psi}}[\mathcal{E}(s)] = \sum_{i=1}^k \tfrac{1}{k} F(\sigma_{i,s}, (V_i \otimes I) \rho (V^\dagger_i \otimes I))    
    \label{eq:tamperability_equation}
\end{equation}
\end{lemma}
\begin{proof}
The joint-state of Eve and Bob, after Eve has applied $V_i$ on her qubit, can be written as $(V_i \otimes I)\ket{\alpha}$.

The probability that the the key-recovery step generates $s$ from this state can be expressed as 
\begin{align*}
    & \left| \braket{s}{U_{\ket{\psi_i}}(V_i \otimes I)|\alpha} \right|^2\\
=   & \left| \braket{\psi_i}{U_s(V_i \otimes I)|\alpha} \right|^2\\
=   & F\left(U_s \ket{\psi_i}\bra{\psi_i}U_s~,~ (V_i \otimes I)\ketbra{\alpha}{\alpha}(V_i^\dagger \otimes I)\right)\\
=   & F(\sigma_{i,s}, (V_i \otimes I)\rho(V_i^\dagger \otimes I)),
\end{align*}
using Eq.~\ref{eqn:recoverability} and conditioned on the nonce being $\ket{\psi}$. The claim follows since the probability of each nonce is $\tfrac{1}{k}$.
\end{proof}

The above lemma allows us to derive a sufficient condition for $\mathcal{E}(s)$ to not hold (i.e., {\em for the scheme to be secure}) that we state in the form of the next lemma. The lemma is obtained by tracing out Eve’s state from Eq.~\ref{eq:tamperability_equation} and using the fact that subsystem fidelity is no less than the fidelity of a larger system. Let $\sigma^B_{i,s}$ denote $\Tr_E{\sigma_{i,s}}$ and $\rho^B$ denote any single-qubit state; define the largest expected fidelity between them as $\mathbb{F}(s)$.
\[
    \mathbb{F}(s) = \max_{\rho^B} \sum_{i=1}^k \tfrac{1}{k} F(\sigma^B_{i,s}, \rho^B).
\]

\begin{lemma} 
The probability that $s$ is recovered is upper bounded as:

\label{lem:tamperability_sufficient_cond}
\begin{equation}
    \Pr_{nonce}[\mathcal{E}(s)] \le \mathbb{F}(s)
\end{equation}
\end{lemma}

For H03-QSS scheme with any set of nonces, $\mathbb{F}(s)$ can be computed by solving the corresponding optimization problem to find the best $\rho^B$ that is independent of $s$. 

{If $s$=00 or 01,} then Eve's attack remains undetected if either {00 or 01} is recovered in Stage-III, and the probability of this happening is upper-bounded by $2\mathbb{F}(s)$. On the other hand, {if $s$=10 or 11}, then Eve's attack remains undetected if exactly $s$ is recovered in Stage-III whose probability is at most $\mathbb{F}(s)$. Thus, a low $\mathbb{F}(s)$ for every $s$ ensures that Eve's attack can be detected almost always.


We actually show that the upper bound in Lemma~\ref{lem:tamperability_necessary_cond} is tight, i.e., we show how Eve can launch an attack such that she manages to have $s$ recovered in the Secret Recovery Stage with the highest possible probability.

\begin{lemma}
\label{lem:tamperability_necessary_cond}
{Let $s$ be some secret. Eve can select an $\ket{\alpha}$ and a set of operators $\{V_1, V_2, \ldots, V_k\}$ such that 
$$ \Pr_{\ket{\psi}}[\mathcal{E}(s)] = \mathbb{F}(s).$$}
\end{lemma}

The proof of this lemma is based on Uhlmann's theorem and is constructive; i.e., the state $\ket{\alpha}$ and the operators $\{V_1, \ldots, V_k\}$ can be explicitly constructed following the steps of the proof of Uhlmann's theorem.


Let us first state the version of Uhlmann's theorem that we use; it a slight variation of the version given by Nielsen and Chuang~\cite{Nielsen_Chuang_2010}.

\begin{theorem}[Uhlmann's theorem]
    Let $\alpha$ and $\beta$ be two states over some Hilbert space $\mathcal{H}^B$. Let $\mathcal{H}_A$ be the Hilbert space of a reference system that is a copy of $\mathcal{H}_B$. 
    Let $\ket{b}$ be any purification of $\beta$ in $\mathcal{H}_A \otimes \mathcal{H}_B$. Then, 
    one can construct a purification $\ket{a} \in \mathcal{H}_A \otimes \mathcal{H}_B$ of $\alpha$ 
    such that the following holds.
    $$F(\alpha, \beta) = F(\ket{a}\bra{a}, \ket{b}\bra{b})$$
\end{theorem}
The proof of Lemma~\ref{lem:tamperability_necessary_cond} is below.
 \begin{proof}
We can directly invoke Uhlmann's theorem to prove the above lemma.
Let $\rho=\ket{\alpha}\bra{\alpha}$ be any purification of $\rho^B$ (thus, $\rho$ is independent of $i,s$); for example, we can use the canonical purification of $\rho^B$.

We immediately get the following.
$$F(\sigma^B_{i,s}, \rho^B) = F(\sigma^*_{i,s}, \rho),$$
where $\sigma^*_{i,s}$ is some purification of $\sigma^B_{i,s}$.

Furthermore, since all purification are equivalent with respect to unitary operations on the reference system, and $\sigma_{i,s}=U_s\ket{\psi_i}\bra{\psi_i}U_s$ is a purification of $\sigma^B_{i,s}$, there exists a unitary operator, say $U_{i,s}$, such that

\begin{align*}
\sigma^{*}_{i,s} &= (U_{i,s} \otimes I)\,\sigma_{i,s}\,(U^\dagger_{i,s} \otimes I) ,\\
F(\sigma^B_{i,s}, \rho^B) 
&= F\!\left((U_{i,s} \otimes I)\,\sigma_{i,s}\,(U^\dagger_{i,s} \otimes I),\, \rho \right) \nonumber \\
&= F\!\left(\sigma_{i,s},\, (U^\dagger_{i,s} \otimes I)\,\rho\,(U_{i,s} \otimes I)\right) .
\end{align*}

\begin{align*}
   \sum_{i=1}^{k} F(\sigma^B_{i,s}, \rho^B) &= \sum_{i=1}^{k} F(\sigma_{i,s}, (U^\dagger_{i,s} \otimes I)\rho (U_{i,s} \otimes I)) \\
   \sum_{i=1}^{k} \frac{1}{k}F(\sigma^B_{i,s}, \rho^B) &= \sum_{i=1}^{k} \frac{1}{k} F(\sigma_{i,s}, (U^\dagger_{i,s} \otimes I)\rho (U_{i,s} \otimes I))\\
   &= \Pr_{\ket{\psi}}[{\mathcal{E}}(s)] \quad \quad (\text{from}     \; \; Eq.~\eqref{eq:tamperability_equation})
\end{align*}
Now $\rho^B$ is any purification for Uhlmann's theorem, hence 
$$\max_{\rho^{B}}\sum_{i=1}^{k} \frac{1}{k}F(\sigma^B_{i,s}, \rho^B) = \Pr_{\ket{\psi}}[\mathcal{E}(s)].$$

\end{proof}


We now state the main theorem of this section that simply combines the main results proved earlier.

\begin{theorem}
    For any fixed $s$, Eve can always launch an intercept-fake-resend attack such that the parties generate $s$ in the Secret Recovery stage with probability $\mathbb{F}(s)$, taken over the choice of the nonce, and this probability cannot be improved any further.
\end{theorem}

Thus, a high-value of $\mathbb{F}(s)$, e.g., $\mathbb{F}(s)=1$, implies that Eve can always evade detection if the secret chosen by the dealer is $s$.

\section{A Limit on One-shot Security}
\label{sec:limit-one-shot}
In this section we discuss a few results that highlight the underlying technical limitations of H03-QSS, even under the best possible choice of nonces.

\subsection{Security against Intercept-fake-resend Attack}\label{subsec:no-attack-not-possible}

In this section we prove, contrary to our initial belief, that Eve always has a chance of avoiding detection after launching an intercept-fake-resend attack for any set of nonces.

First, we prove that average fidelity of a set of single-qubit (possibly mixed) states with an optimally chosen single-qubit (possibly mixed) state is at at least $\tfrac{1}{2}$.


\begin{lemma}
{	Let $\beta_1, \ldots \beta_k$ denote any $k$ single-qubit (possibly mixed) states. Then, there exists a single-qubit state $\gamma$ (possible mixed) such that the average fidelity of $\beta_i$ from $\gamma$ is at least $\tfrac{1}{2}$.} 
    $$\max_{\rho_{_\gamma}}\sum_{i=1}^{k} \frac{1}{k}F(\rho_{_{\beta_i}} , \rho_{_{\gamma}}) \geq \frac{1}{2}$$
    \label{lem:lower-bound-average-fidelity}
\end{lemma}

\begin{proof}
The fidelity between two single qubit states $\rho_{\beta}$ and $\rho_{\gamma}$ can be written as~\cite{Jozsa01121994}
\begin{align*}
F(\rho_\beta,\rho_\gamma)
  & = \frac12\!\left[1 + \vec{\beta}\cdot \vec{\gamma}
      + \sqrt{(1-\lVert \vec{\beta}\rVert^{2})(1-\lVert \vec{\gamma}\rVert^{2})}\right]\\
  & \ge \frac12\!\left[1 + \vec{\beta}\cdot \vec{\gamma}\right].
\end{align*}
where, $\vec{\beta}$ and $\vec{\gamma}$ denote the Bloch vectors of $\rho_B$ and $\rho_\gamma$, respectively.
Expanding 
$\vec{\beta}\cdot\vec{\gamma} 
   = \frac12 \Big( \lVert \vec{\beta}\rVert^{2} + \lVert \vec{\gamma}\rVert^{2} 
    - \lVert \vec{\beta}-\vec{\gamma}\rVert^{2} \Big)$ allows us to lower bound the above fidelity as:  
\[
F(\rho_\beta,\rho_\gamma)
  \ge \frac12 + \frac14\!\left[ \lVert \vec{\beta}\rVert^{2}
 + \lVert \vec{\gamma}\rVert^{2} - \lVert \vec{\beta}-\vec{\gamma}\rVert^{2} \right].
\]

We will use this bound on the sum of the fidelities in the statement of the lemma. Denote the Bloch vector of $\rho_{\beta_i}$ by $\vec{\beta_i}$.



\begin{align*}
\max_{\rho_{_\gamma}} & \sum_{i=1}^k 
F(\rho_{\beta_i},\rho_\gamma)
\ge\\
& \frac{k}{2}
+ \frac{1}{4}\max_{\gamma}
\left[
\sum_{i=1}^k \lVert \vec{\beta}_i\rVert^{2}
+ k \lVert \vec{\gamma}\rVert^{2}
- \sum_{i=1}^k\lVert \vec{\beta}_i - \vec{\gamma}\rVert^{2}
\right].  
\end{align*}


Let's expand the last sum in the above expression.
\[
\sum_{i=1}^k \lVert \vec{\beta}_i - \vec{\gamma}\rVert^{2}
= 
\sum_{i=1}^k \lVert \vec{\beta}_i\rVert^{2}
+ k \lVert \vec{\gamma}\rVert^{2}
- 2\left(\sum_{i=1}^k \vec{\beta}_i\right)\cdot \vec{\gamma}.
\]

Substituting gives us
\begin{equation}
\max_{\rho_{_\gamma}} \sum_{i=1}^k 
F(\rho_{\beta_i},\rho_\gamma) \ge
\frac{k}{2} + \frac{k}{2} \max_\gamma \vec{\beta} \cdot \vec{\gamma},
\label{eq:lemma7_1}
\end{equation}
where, $\bar{\beta} = \frac{1}{k}\sum_{i=1}^k \vec{\beta}_i$. The dot product is maximized when $\vec{\gamma}$ is parallel to $\bar{\beta}$, i.e.,  choose $\vec{\gamma}_{\max} = c \bar{\beta}$ for some positive $c$.



Since $c$ and $k$ are both constants, we can rewrite Eq.~\ref{eq:lemma7_1}:
\[
\max_{\rho_{_\gamma}} \sum_{i=1}^k 
F(\rho_{\beta_i},\rho_\gamma) \ge \frac{k}{2} + \frac{ck}{2}\lVert\bar{\beta}\rVert^{2} 
\ge \frac{k}{2}
\]


Hence, the average fidelity satisfies
\[
\max_{\rho_{_\gamma}}
\frac{1}{k}\sum_{i=1}^k F(\rho_{\beta_i},\rho_\gamma)
\ge \frac12.
\]
\end{proof}

Recall that Eve first generates the dealer's secret $s$ in Stage-II using the knowledge of $\ket{\psi_i}$, and then manipulates the joint-state $\ket{\alpha}$ accordingly. 
The above lemma essentially lower bounds $\mathbb{F}(s) \ge \tfrac{1}{2}$, for any $s$ Eve desires, leading to the following corollary.

\begin{corollary}\label{cor:imr-always-exists}
    Irrespective of the number of nonces, for any $s$, Eve can suitably tamper the protocol in Stage-I such that exactly $\ket{s}$ is generated in the key-recovery stage with probability at least \textonehalf.
\end{corollary}

The corollary does not quite imply that Eve can manipulate the protocol to have the $s$ chosen by the dealer be generated in Stage-III. This is due to the reason that $\gamma$ in the above lemma (whose purification would be the state $\ket{\alpha}$ shared by Eve) may be different for different sets of $\{ \beta_i \}_{i=1}^k$.

Nevertheless, we show how to lower bound the success probability of Eve.
As per Lemma~\ref{lem:lower-bound-average-fidelity}, {$\mathbb{F}(01) \ge \tfrac{1}{2}$}. Further, using Lemma~\ref{lem:tamperability_necessary_cond}, Eve can determine $\ket{\alpha}$ and a set of operators $\{V_i\}_{i=1}^k$ such that the parties can {recover} 01 in Stage-III with probability at least $\tfrac{1}{2}$.

Now, here is one strategy that Eve can adopt:
Eve selects $\ket{\alpha}$ and a set of operators $\{V_i\}_{i=1}^k$ {such that the  string} ``01'' is generated in Stage-III, regardless of the announced nonce and the $s$ she computes in Stage-II.

Recall that, in the $\mathtt{SECRET}$ mode, attack is not detected if 01 is generated in Stage-III. Since the probability of choosing this mode is exactly \textonehalf, Eve can succeed with probability at least $(1/2) \cdot (1/2) = 1/4$. We summarise this as the following theorem.

\begin{theorem}
	\label{thm:lower-bound-protocol}
	{For any set of nonces, Eve can always launch an intercept-fake-resend attack whose probability of detection is at most $\tfrac{3}{4}$.} 
\end{theorem}

\subsection{Secrecy Implies Intercept-fake-resend Attack}
\label{subsec:secrecy-implies-imr}

In Section~\ref{subsec:no-attack-not-possible} we analysed the security of the protocol against the \emph{intercept-fake-resend} attack. In particular, Corollary~\ref{cor:imr-always-exists} established that Eve can always ensure that a particular $s$ is generated in Stage-III, thus leading to a successful attack with probability at least $\tfrac{1}{4}$. This begs the question whether the dealer can cap the probability of attack to $\tfrac{1}{4}$ and no more. We show that this is not possible unless the dealer wants to lower the expectation of secrecy. 

We essentially prove a trade-off between secrecy and protection against intercept-fake-resend attack assuming that the dealer designs a perfectly recoverable scheme.



Consider a set of nonces such that the secret is perfectly recoverable (Lemma~\ref{lemma:recoverability_condition}). We start by characterising the individual shares of Eve and Bob for this scenario.
\begin{lemma}
    Recoverability implies that then the reduced density operators of Eve and Bob must lie on the $XY$-plane of the Bloch sphere.
    {\label{lem: recoverability imples XY-plane}}
\end{lemma}
\begin{proof}
    Let $\ket{\psi_i}$ denote any nonce that satisfies recoverability. Then  $|\langle s|\psi_i \rangle| = \tfrac{1}{2}$ for all $s \in \{00,01,10,11\}$. Therefore, $\ket{\psi_{i,s}} = U_s\ket{\psi}$ can be written as  
\begin{equation}
U_s\ket{\psi} = \tfrac{1}{2}[\ket{00} + e^{i\alpha_1}\ket{01} + e^{i\alpha_2}\ket{10} + e^{i\alpha_3}\ket{11}].\label{eq:lemma_8_1}
\end{equation}

Suppose the first qubit is sent to Eve and the second one to Bob. Let's analyse $\sigma^B_{i,s} = \Tr_E{\ketbra{\psi_{i,s}}}$ that represents the state of Bob's share; that of Eve's share has a similar analysis.

It is straightforward to show the following from Eq.~\ref{eq:lemma_8_1}:
$$\bra{0}\sigma^B_{i,s} \ket{0} = \bra{1}\sigma^B_{i,s} \ket{1} = \tfrac{1}{2}$$
This implies that $\sigma^B_{i,s}$, when represented in terms of its Bloch vectors as $\sigma^B_{i,s} = \frac{I}{2} + \tfrac{1}{2}[r_X \sigma_X + r_Y \sigma_Y + r_Z \sigma_Z]$, must have $r_Z = 0$, i.e., $\sigma^B_{i,s}$ must lie on the $XY$ plane.

\end{proof}

The next lemma characterises $\sigma_{i,s}$.


\begin{lemma}
    Recoverability implies that $\ket{\psi_{i,s}}$ has one of the following structures.
    \begin{itemize}
        \item  $\ket{\psi_{i,s}}$ is a bipartite state $\ket{E_{i,s}}\ket{B_{i,s}}$ in which both $\ket{E_{i,s}}$ and $\ket{B_{i,s}}$ lie on the $XY$-plane. 
        \item $\ket{\psi_{i,s}}$ is an equal superposition of 
$\ket{E_{i,s}}\ket{B_{i,s}}$ and $\ket{E_{i,s}^\perp}\ket{B_{i,s}^\perp}$
in which $\ket{E_{i,s}}$ and $\ket{E_{i,s}^\perp}$ lie on the $Z$-axis but $\ket{B_{i,s}}$ and $\ket{B_{i,s}^\perp}$ lie on the $XY$-plane.
    \end{itemize}
    {\label{lem: recoverability:c1 c2}}
\end{lemma}
\begin{proof}
We will use the fact that $\ket{\psi_{i,s}}$ is a purification of $\sigma^B_{i,s}$ and $\sigma^E_{i,s}$ both of which must be on the $XY$-plane of the Bloch sphere.

Any mixed state on the on the $XY$-plane can be expressed as
$$c_1 \ket{e_1}\bra{e_1} + c_2 \ket{e_1^{\perp}}\bra{e_1^{\perp}},$$ 
where $c_1,c_2$ are (real) positive constants adding up to 1, $\ket{e_1} = \frac{1}{\sqrt{2}}(\ket{0}+ e^{i\phi_1}\ket{1})$, and $\ket{e_1^\perp}$ is orthogonal to $\ket{e_1}$.

This allows us to represent $\ket{\psi_{i,s}}$ using Schmidt decomposition:
 $$ \ket{\psi_{i,s}} = \sqrt{c_1} \ket{E_{i,s}}\ket{B_{i,s}} + \sqrt{c_2}\ket{E_{i,s}^{\perp}} \ket{B_{i,s}^{\perp}},$$
where,
$$\ket{B_{i,s}} = \tfrac{1}{\sqrt{2}}(\ket{0} + e^{i \phi_i^s} \ket{1})~ \text{and}\quad \ket{E_{i,s}}=\cos\frac{\theta}{2}\ket{0}+ e^{i\mu_i^s}\sin\frac{\theta}{2}\ket{1}$$
are two single-qubit pure states (here, $\phi_i, \mu_i \in [0,2\pi]$ and $\theta \in [0,\pi]$).

Let's write amplitudes of $\ket{00}, \ket{01}, \ket{10}, \ket{11}$ in  $\ket{\psi_{i,s}}$.
\begin{align*}
& \text{amplitude of }\ket{00} = \frac{1}{\sqrt{2}}\left[\sqrt{c_1}\cos\frac{\theta}{2} - \sqrt{c_2}\sin\frac{\theta}{2}\right]\\
& \text{amplitude of }\ket{01} = \frac{1}{\sqrt{2}}\left[\sqrt{c_1}\cos\frac{\theta}{2} + \sqrt{c_2}\sin\frac{\theta}{2}\right]e^{i \phi_i^s}\\
& \text{amplitude of }\ket{10} = \frac{1}{\sqrt{2}}\left[\sqrt{c_1}\sin\frac{\theta}{2} + \sqrt{c_2}\cos\frac{\theta}{2}\right]e^{i \mu_i^s}\\
& \text{amplitude of }\ket{11} = \frac{1}{\sqrt{2}}\left[\sqrt{c_1}\sin\frac{\theta}{2} - \sqrt{c_2}\cos\frac{\theta}{2}\right]e^{i (\phi_i^s + \mu_i^s)}    
\end{align*}

We know from Eq.~\ref{eq:lemma_8_1} that recoverability also implies that the above amplitudes have their real parts as $\tfrac{1}{2}$.
 \begin{align}
     \left|\sqrt{c_1}\,\cos\frac{\theta}{2} - \sqrt{c_2}\,\sin\frac{\theta}{2}\right| = \frac{1}{\sqrt{2}} \tag{$P_1$}\\
     \left|\sqrt{c_1}\cos\frac{\theta}{2} + \sqrt{c_2}\sin\frac{\theta}{2}\right| = \frac{1}{\sqrt{2}} \tag{$P_2$}\\
     \left| \sqrt{c_1}\sin\frac{\theta}{2} + \sqrt{c_2}\cos\frac{\theta}{2}\right| = \frac{1}{\sqrt{2}} \tag{$P_3$}\\
     \left| \sqrt{c_1}\sin\frac{\theta}{2} - \sqrt{c_2}\cos\frac{\theta}{2}\right| = \frac{1}{\sqrt{2}}  \tag{$P_4$}
 \end{align}
 
From $P_3$ and {$P_4$}, we get these.
 \begin{align}
     c_1\, \sin^2{\frac{\theta}{2}} +  c_2\, \cos^2{\frac{\theta}{2}} & = \frac{1}{2}  \tag{$P_5$}\\
    2\sqrt{c_1}\sqrt{c_2}\,\sin{{\theta}} & =0  \tag{$P_6$}     
 \end{align}

Similarly, from $P_1$ and $P_2$, we get
\begin{align}
c_1 \cos^2 \frac{\theta}{2} + c_2 \sin^2 \frac{\theta}{2} = \frac{1}{2} \tag{$P_7$}
\end{align}

Equations $P_5,P_6$ and $P_7$ imply that one of these hold.
\begin{itemize}
    \item $c_1=0$, implying that $c_2=1$ and $\theta = \tfrac{\pi}{2}$.
    \item $c_1=1$, implying that $c_2=0$ and $\theta = \tfrac{\pi}{2}$.
    \item $\theta=0$ or $\pi$, implying that $c_1=c_2=\tfrac{1}{2}$.
\end{itemize}
For the first two scenarios, $\ket{\psi_{i,s}}$ is a bipartite state $\ket{E_{i,s}}\ket{B_{i,s}}$. By Lemma~\ref{lem: recoverability imples XY-plane}, both $\ket{E_{i,s}}$ and $\ket{B_{i,s}}$ lie on the $XY$-plane.

For the third, $\ket{\psi_{i,s}}$ is an equal superposition of 
$\ket{E_{i,s}}\ket{B_{i,s}}$ and $\ket{E_{i,s}^\perp}\ket{B_{i,s}^\perp}$. Since $\theta=0$ or $\pi$, $\ket{E_{i,s}}$ lie on the $Z$-axis and $\ket{B_{i,s}}$ lie on the $XY$-plane. For any state on the $XY$-plane, its orthogonal state also lies on the $XY$-plane, and for any state on the $Z$-axis, its orthogonal state lies on the $Z$-axis. Thus, both 
$\ket{E_{i,s}}$ and $\ket{E_{i,s}^\perp}$ lie on the $Z$-axis and both $\ket{B_{i,s}}$ and $\ket{B_{i,s}^\perp}$ lie on the $XY$-plane.

The required claims about the Schmidt coefficients are obtained.
\end{proof}

We are now ready to state and prove our theorem capturing the trade-off between secrecy and intercept-fake-resend protection.

\begin{theorem}\label{theorem:secrecy-implies-tamperability}
    Consider an H03-QSS scheme that is perfectly recoverable. At least one of the following is always possible.
    \begin{enumerate}
        \item Eve can mount an intercept-fake-resend attack that cannot be detected at all.
        \item Secrecy is violated for one of the parties (among Eve and Bob), i.e., one of the parties can determine the secret with probability 1 without launching any interception attack.
    \end{enumerate}
\end{theorem}



\begin{proof}
Consider an H03-QSS scheme with a set of nonces
$\{\ket{\psi_i}\}_{i\in J}$.

Without loss of generality, we focus on the nonce $\ket{\psi_1}$, since the analysis for the other states $\ket{\psi_i}$ proceeds in an identical manner. 
In Lemma~\ref{lem: recoverability:c1 c2}, we proved that $\ket{\psi_{1,01}}$ is either a bipartite state or an equal superposition. Next, we analyse these two possibilities separately.

Let's begin by assuming that $\ket{\psi_{1,01}}$ is a bipartite state $\ket{E_{1,01}}\ket{B_{1,01}}$ in which both states lie on the $XY$-plane. Thus, we can represent 
\begin{align*}
\ket{\psi_{1,01}} & = \ket{E_{1,01}} \otimes \ket{B_{1,01}} \\
    & = \tfrac{1}{\sqrt{2}}(\ket{0} + e^{i\phi_1} \ket{1})\otimes \tfrac{1}{\sqrt{2}}(\ket{0} + e^{i\mu_1} \ket{1})    \\
    & = \tfrac{1}{2}[ \ket{00} + e^{i\mu_1} \ket{01} + e^{i \phi_1} \ket{10} + e^{i(\phi_1 + \mu_1)} \ket{11}].
\end{align*}

Observe that $\ket{\psi_{1,10}} = U_{10} \ket{\psi_1} = U_{10} U_{01} \ket{\psi_{1,01}}$.
\begin{align*}
\ket{\psi_{1,10}}  & = \tfrac{1}{2}[ \ket{00} - e^{i\mu_1} \ket{01} - e^{i \phi_1} \ket{10} + e^{i(\phi_1 + \mu_1)} \ket{11}]\\
    & = \tfrac{1}{\sqrt{2}}(\ket{0} - e^{i\phi_1} \ket{1})\otimes \tfrac{1}{\sqrt{2}}(\ket{0} - e^{i\mu_1} \ket{1})    \\
    & = \text{(denoted)~}\ket{E_{1,01}^\perp} \otimes \ket{B_{1,01}^\perp} \\
\end{align*}

So, the states Eve receives for $s=01$ and $s=10$ are pure and orthogonal to each other implying that $\| \sigma^E_{1,01} - \sigma^E_{1,10} \|_1=2$. Since Eve can distinguish between these states with probability 1, she violates the requirement of best possible secrecy at her end (see Lemma~\ref{lemma:eve-guessing-secret}).

A similar attack on secrecy at Bob's end also exists. 
Thus, should we care for perfect secrecy, {\em all the nonces} should be chosen such that for all $i$, $\ket{\psi_{i,01}}$ is an equal superposition state of the form specified in Lemma~\ref{lem: recoverability:c1 c2}, i.e.,
$$\ket{\psi_{i,01}} = \tfrac{1}{\sqrt{2}} \ket{E_{i,01}} \ket{B_{i,01}} + 
\tfrac{1}{\sqrt{2}} \ket{E_{i,10}^\perp} \ket{B_{i,10}^\perp} $$
in which $\ket{E_{i,10}}$ and $\ket{E_{i,10}^\perp}$ lie on the $XY$-axis and $\ket{B_{i,10}}$ and $\ket{B_{i,10}^\perp}$ lie on the $Z$-axis.

Let's compute Bob's state for any nonce $\ket{\psi_i}$ and $s$.

$$\sigma^B_{i,s} = \Tr_E{\ket{\psi_{i,s}}\bra{\psi_{i,s}}} = \tfrac{1}{2} \ketbra{B_{i,s}} + \tfrac{1}{2} \ketbra{B_{i,s}^\perp} = I/2$$


When the reduced density operator of Bob equals $\frac{I}{2}$ for all $i$, it is known that Eve can launch an intercept-fake-resend attack that cannot be detected with non-zero probability; in fact, the details of such an attack has already been discussed in Section~\ref{subsec:hao}. 


Thus, preventing both the scenarios in the statement of the theorem is not possible.

\end{proof}

\section{Analysing Original H03-QSS scheme}
\label{sec:analysis_original_Hsu}
In this section we analyse the vulnerabilities of the original H03-QSS scheme using our characterization.



The original scheme uses 16 nonces formed by taking all possible combinations of two qubits from the set \(T=\{\ket{+}, \ket{-}, \ket{+i}, \ket{-i}\}\). 
These nonces can be shown to satisfy recoverability, has perfect secrecy, and ensures protection against the intercept-measure-resend attack.

It is straightforward to verify that for any nonce $\ket{\psi}$ and any $s \in \{ 00,01,10,11\}$, $|\braket{s}{\psi}| = \tfrac{1}{2}$ (see Table~\ref{tab:hsu_nonces}), and thus, all its secrets are recoverable as per Lemma~\ref{lemma:recoverability_condition}.

To show that secrecy is preserved, the following lemma will be useful. It can be directly verified from the nonces; the complete table of $\{\sigma_{i,s}\}_{i,s}$ is presented in Appendix~\ref{appendix:hsu_sigma_table}.

\begin{table}[]
    \centering
    \renewcommand{\arraystretch}{1.5}
    \begin{tabular}{|lll|}
    \hline
    $\ket{\psi_{1}}=$ & $\ket{+}\ket{+} =$ & $ \frac{1}{2} [\ket{00} + \ket{01} + \ket{10} + \ket{11}]$ \\
    $\ket{\psi_{2}}=$ & $\ket{-}\ket{-}=$ & $\frac{1}{2} [\ket{00} -  \ket{01} - \ket{10} + \ket{11}]$\\
    $\ket{\psi_{3}}=$ & & $\frac{1}{2} [\ket{00} + \ket{01} - \ket{10} + \ket{11}]$\\
    $\ket{\psi_{4}}=$ & & $\frac{1}{2} [\ket{00} -\ket{01} + \ket{10} + \ket{11}]$\\
    \hline 
    \end{tabular}
    \caption{Nonce states proposed by us.}
    \label{tab:our_nonces}
\end{table}

\begin{lemma}\label{lemma:sigma_B_Hsu}
    For any $i$ and $s$, $\Tr_E{\sigma_{i,s}} = \Tr_B{\sigma_{i,s}} = I/2$.
\end{lemma}

The above lemma immediately implies that the condition in Lemma~\ref{lemma:secrecy} is met, i.e., the original protocol preserves perfect secrecy for both the parties.

To analyse the scheme's protection against an intercept-measure-resend attack, observe that
\begin{align*}
    \sum_i \sigma_{i,01} &= \sum_i \ket{\psi_{i,01}}\bra{\psi_{i,01}} \\
    &= \sum_i U_{01} \ket{\psi_i} \bra{\psi_i} U_{01} \\
    &= U_{01} \left( \sum_{b \in T} \sum_{d \in T} \ketbra{b} \otimes \ketbra{d} \right) U_{01}\\
    &= U_{01} \cdot 4I \cdot U_{10} = 4I.
\end{align*}

A similar analysis shows that $\sum_i \sigma_{i,10}=4I$. 
Thus, according to Lemma~\ref{lemma:imr}, the nonces are able to prevent an intercept-measure-resend attack.

However, it completely fails to protect against the intercept-fake-resend attack. We will show this by computing $\mathbb{F}(s)$ for every $s$. 

We showed in Lemma~\ref{lemma:sigma_B_Hsu} that $\sigma^B_{i,s} = \Tr_E{\sigma_{i,s}} = I/2$ for all nonce $\ket{\psi_i}$ and all secret $s$. 
Now, let's choose any 2-qubit state $\ket{\alpha}$ that is a purification of $I/2$; Hao et al.~\cite{hao2010eavesdropping} chose $\ket{\alpha} = \tfrac{1}{\sqrt{2}}[\ket{01} + \ket{10}]$ but we can chose any other state too, e.g., $\tfrac{1}{\sqrt{2}}[\ket{++} - \ket{--}]$. Clearly, $F(\sigma^B_{i,s}, \ket{\alpha}\bra{\alpha})=1$, and thus $\mathbb{F}(s)=1$ for all $s$. 

Using $\ketbra{\alpha}$ as $\rho^B$ in lemma~\ref{lem:tamperability_necessary_cond}, we can conclude that Eve can always launch an attack by sharing a suitable $\ket{\alpha}$ between her and Bob and later tampering it remotely using an operator from a set $\{V_1, \ldots, V_k\}$.


%
It can be further verified that if $\ket{\alpha}$ is chosen to be $\tfrac{1}{\sqrt{2}}[\ket{01} + \ket{10}]$, then the operators match exactly those proposed by Hao et al.~\cite[Table~2]{hao2010eavesdropping}.

The following theorem summarises the appropriateness of the original nonces.
\begin{theorem}\label{thm:hsu_nonces}
    H03-QSS using the nonces shown in Table~\ref{tab:hsu_nonces} is recoverable, ensures secrecy for both the parties, prevents an intercept-measure-resend attack but fails against an intercept-fake-resend attack, i.e., a Eve can learn the secret after intercepting the qubits {\it en route} without anyone noticing the attack anytime ever after.
\end{theorem}

\section{Fortifying the H03-QSS scheme}
\label{section:analysis_our_scheme}

In this section we propose a new of set of nonces that improves the security of H03-QSS with respect to the intercept-fake-resend attack by slightly compromising on the secrecy of the secrets. To strengthen the scheme, we propose a set of four nonces, denoted $J$, and listed in Table~\ref{tab:our_nonces}. We retained two of the original nonces and added two of our own.


\setlength\intextsep{1pt}
\begin{table}
\begin{center}
\[
\renewcommand{\arraystretch}{1.3}
\begin{array}{|c|r|r|r|r|}
\hline
\mbox{nonces in $J$} & {\ket{\psi_1}} & {\ket{\psi_2}} & \ket{\psi_3} & {\ket{\psi_4}} \\
\hline
s=00 & U_{00}\ket{++} & U_{00} \ket{--} & -\ket{+}\ket{-} & -\ket{-}\ket{+}\\
\hline
s=01 &  U_{01}\ket{++} & U_{01} \ket{--} & \ket{-}\ket{-} & \ket{+}\ket{+} \\
\hline
s=10 &  U_{10}\ket{++} & U_{10} \ket{--} & \ket{+}\ket{+} & \ket{-}\ket{-} \\
\hline
s=11 &  U_{11}\ket{++} & U_{11} \ket{--} & \ket{-}\ket{+} & \ket{+}\ket{-}  \\
\hline
\end{array}\] \end{center}
    \caption{The table shows the states $\ket{\psi_{i,s}} = U_s\ket{\psi_i}$ for every nonce $\ket{\psi_i} \in J$ and all secret $s$.}
    \label{tab:all_shares}
\end{table}

\subsection{Recoverability}
\label{subsec:new_nonce_recoverability}
It is easy to verify from Table~\ref{tab:our_nonces} that $|\braket{s}{\psi}| = \tfrac{1}{2}$ for all $s \in \{00,01,10,11\}$ and $\ket{\psi}$. Thus, by Lemma~\ref{lemma:recoverability_condition}, the QSS scheme using $J$ as nonces is perfectly recoverable.

\subsection{Secrecy}\label{subsec:new_nonce_secrecy}

Let's discuss the secrecy of Bob first. 

We can use Table~\ref{tab:all_shares_Bob_state} to calculate that

\[
\sum_{i=1}^4 \| \sigma^B_{i,01} - \sigma^B_{i,10} \|_1 = 2 \| \ketbra{+} - \ketbra{-} \|_1 = 4.
\]

Based on this and by Lemma~\ref{lemma:secrecy}, secrecy of Bob is $\tfrac{3}{8}$ which is quite close to the optimal value of \textonehalf. Thus, Bob can independently guess the secret with expected probability at most $\tfrac{5}{8}$ where the expectation is taken over the randomly chosen nonce.

A similar analysis for Eve also shows that Eve too can guess the secret with expected probability at most $\tfrac{5}{8}$.

\subsection{Intercept-measure-resend}\label{subsec:new_nonce_imr}
Next we will show that the new nonces satisfy the conditions of Lemma~\ref{lemma:imr}. First, observe in Table~\ref{tab:all_shares} that
$$U_{01}\ket{--} = U_{10} \ket{++}\quad\text{and}\quad U_{10} \ket{--} = U_{01} \ket{++}.$$

Thus, both $\sum_{i=1}^4 \sigma_{i,01}$ and $\sum_{i=1}^4 \sigma_{i,10}$ are equal mixtures of these four states: 
$$U_{01}\ket{++}, U_{01} \ket{--}, \ket{--}, \ket{++}.$$

This implies that the condition in Lemma~\ref{lemma:imr} holds, and thus, the new nonces provide protection against any intercept-measure-resend attack.

\subsection{Intercept-fake-resend Prevention}
\label{subsec:new_nonce_tamperability}




\begin{table}
\begin{center}
\[
\renewcommand{\arraystretch}{2}
\begin{array}{|c|r|r|r|r|}
\hline
\mbox{nonces in $J$} & \ket{\psi_1} & \ket{\psi_2} & \ket{\psi_3} & \ket{\psi_4} \\
\hline
s=00 & I/2 & I/2 & \ket{-}\bra{-} & \ket{+}\bra{+} \\
\hline
s=01 & I/2 & I/2 & \ket{-}\bra{-} & \ket{+}\bra{+} \\
\hline
s=10 & I/2 & I/2 & \ket{+}\bra{+} & \ket{-}\bra{-} \\
\hline
s=11 & I/2 & I/2 & \ket{+}\bra{+} & \ket{-}\bra{-}  \\
\hline
\end{array}
\]
\end{center}
    \caption{States $\sigma^B_{i,s}$ for every nonce $\ket{\psi_i}$ and every $s$.}
    \label{tab:all_shares_Bob_state}
\end{table}

\begin{table}
\begin{center}
\[
\renewcommand{\arraystretch}{2}
\begin{array}{|c|r|r|r|r|}
\hline
\mbox{nonces in $J$} & \ket{\psi_1} & \ket{\psi_2} & \ket{\psi_3} & \ket{\psi_4} \\
\hline
s=00 & I/2 & I/2 & \ket{+}\bra{+} & \ket{-}\bra{-} \\
\hline
s=01 & I/2 & I/2 & \ket{-}\bra{-} & \ket{+}\bra{+} \\
\hline
s=10 & I/2 & I/2 & \ket{+}\bra{+} & \ket{-}\bra{-} \\
\hline
s=11 & I/2 & I/2 & \ket{-}\bra{-} & \ket{+}\bra{+}  \\
\hline
\end{array}
\]
\end{center}
    \caption{States $\sigma^E_{i,s}$ for every nonce $\ket{\psi_i}$ and every $s$.}
    \label{tab:all_shares_Eve_state}
\end{table}

We will start by computing $\mathbb{F}(s)$ as we did in Section~\ref{sec:analysis_original_Hsu}. Fix any $s$, and let $\rho^B$ be any single-qubit (possibly mixed) state. 
From Table~\ref{tab:all_shares_Bob_state} we can observe that the multiset $\{\sigma^B_{i,s} ~:~ i \in \{1,2,3,4\} \}$ is exactly $\{ I/2, I/2,\ket{+}\bra{+}, \ket{-}\bra{-} \}$ two of which are pure states. We will use the fact that for pure $\sigma^B_{i,s}$, $F(\sigma^B_{s,i},\rho^B) = \Tr{\sigma^B_{s,i}\rho^B}$.

\begin{align*}
&\sum_{i\in\{1,2,3,4\}} F(\sigma^B_{s,i},\rho^B) \\
&= 2F(I/2, \rho^B) + \Tr{\ketbra{+}\rho^B} + \Tr{\ketbra{-}\rho^B} \\
&\le 2 + \Tr{I \cdot \rho^B} = 3.
\end{align*}

So, we get that 
$$\mathbb{F}(s) = \tfrac{1}{4} \sum_{i\in\{1,2,3,4\}} F(\sigma^B_{s,i},\rho^B) \le \tfrac{3}{4}.$$

Therefore, Eve won't be able to force generation of a {\em particular $s$} in Stage-III with probability more than $1/4$. 

Recall that, in the $\mathtt{DETECT}$ mode, the attack is detected if the exact $s$ chosen by the dealer ($s \in \{00, 11\}$) is not generated; we show above that this happens with probability at least $1/4$.
On the other hand, in the $\mathtt{SECRET}$ mode, the attack is \emph{not} detected if either 01 or 10 is generated, 
and Eve may be succeed in that with a very high probability.

Since the dealer choses each mode with equal probabilities, the overall probability that Eve's attack can be detected is at least $\tfrac{1}{8}$.





We state the overall behaviour of our nonces in the following lemma.
\begin{lemma}
    The probability of detecting an intercept-fake-resend attack using our nonces is at least $\tfrac{1}{8}$.
\end{lemma}

While this may not meet our general expectations, recall that the probability is zero in the original scheme.

\subsection{Discussion}

The overall behaviour of our nonces with respect to security and correctness is given below.

\begin{theorem}
    H03-QSS using the nonces shown in Table~\ref{tab:our_nonces} is recoverable, and it prevents any intercept-measure-resend attack. Further, no honest party can guess the secret on its own with probability more than $\tfrac{5}{8}$ and any attempt by the dishonest party to learn the secret using an intercept-fake-resend attack will be detected with probability at least $\tfrac{1}{8}$.
    \label{thm:our_nonces}
\end{theorem}

It is important to contrast this with Theorem~\ref{thm:hsu_nonces} and observe that while the latter does not allow an honest party to guess the secret, it allows a dishonest party to learn the secret in a completely covert manner. An additional consideration is the fact that in this framework the dishonest party {\em always} has a reasonable chance ($\tfrac{1}{4}$) of remaining undetected after mounting an attack (Theorem~\ref{thm:lower-bound-protocol}), no matter what nonces are used.

\vspace{4mm}

\section{Conclusion}
In this work we have analysed the conditions of correctness and security (against different types of attacks) on a one-shot supposedly-perfect quantum secret sharing scheme~\cite{hsu-scheme} denoted H03-QSS. Even though the author of the scheme included various heuristics to prevent the attacks and explained how various attacks can be prevented, his analysis was done by considering a few typical operations an adversary could launch. A subsequent work by Hao et al.~\cite{hao2010eavesdropping} proved him wrong. 

The benefit and importance of mathematical models of attacks and provable claims of security is, thus, super-critical for security protocols.
Riding on the above sentiments, we present a formal model of eavesdropping attack for H03-QSS and explain how its security against this attack can be improved as compared to the original version. We analysed two-types of eavesdropping attacks, one in which Eve measures the intercepted state (and thus destroys it), and in another, retains it until the dealer announces a nonce. However, we show that it is impossible to design a perfectly one-shot secret sharing scheme in this framework. Thus, the quest of such a scheme, of-course in a different framework, continues.



It is easy to workaround the vulnerability arising from the original nonces by running the protocol for multiple rounds, involving a few decoy ones, and running statistical tests~\cite{hao2010eavesdropping}. This of course destroys the one-shot nature.

In our opinion, all the limitations of the H03-QSS framework arises from the requirement of recoverability (see Section~\ref{subsec:recoverability}). Recoverability imposes that the nonces have equiprobable amplitudes, and this forces the states of the shares obtained by the individual parties to take specific symmetric forms. The lower bound proofs in Section~\ref{sec:limit-one-shot} heavily rely on the symmetries.
The conditions for recoverability, in turn, are derived from the reflection operations in the Grover's iterator (Eq.~\ref{eq: grover_reflection}). Thus, a natural direction towards designing a more resilient one-shot framework is to to explore reconstruction mechanisms beyond these operations.


\bibliographystyle{apsrev4-2}
\bibliography{refs}

\clearpage

\appendix
\section{Appendix: Proof of Lemma~\ref{lemma:recoverability_condition}}
\label{appendix:lemma_recoverability_proof}

\begin{proof}

Let's first extend $\{\ket{s}\}$ to some basis $\{ \ket{b_1}, \ket{b_2}, \ldots \ket{b_n}\}$. Without loss of generality, let $\ket{b_1} = \ket{s}$. Now we can represent any nonce $\ket{\psi}$ in this basis.
\begin{center}
$\ket{\psi} = \sum_{i=1}^{N} \alpha_i \ket{b_i}$
\end{center}

The shared secret in this case would be $$
\ket{\psi_{i,s}} = U_{\ket{s}} \ket{\psi} = \sum_{\substack{i=2}}^{N} \alpha_i \ket{b_i} - \alpha_1 \ket{b_1}
$$
\begin{widetext}
Now,
\begin{align*}
U_{\ket{\psi}} U_{\ket{s}} \ket{\psi} &= -\left( 2 \ket{\psi} \bra{\psi} - I \right) \left[ \sum_{\substack{i=2}}^{N} \alpha_i \ket{b_i} - \alpha_1 \ket{b_1} \right] \\
&= -2 \ket{\psi} \left\langle \psi \middle| \sum_i \alpha_i \ket{b_i} - \alpha_1 \ket{b_1} \right\rangle 
+ \left[ \sum_{\substack{i=2}}^{N} \alpha_i \ket{b_i} - \alpha_1 \ket{b_1} \right]\\
&= -2 \ket{\psi} \left[ \sum_{\substack{i=2}}^{N} |\alpha_i|^2 - |\alpha_1|^2 \right] 
+ \left[ \sum_{\substack{i=2}}^{N} \alpha_i \ket{b_i} - \alpha_1 \ket{b_1} \right]\\
&= -2 \ket{\psi} \left[ \sum_{\substack{i=2}}^{N} |\alpha_i|^2 - |\alpha_1|^2 \right] 
+ \left[ \sum_{\substack{i=2}}^{N} \alpha_i \ket{b_i} - \alpha_1 \ket{b_1} \right]\\
&= -2 \ket{\psi} \left[ 1 - 2|\alpha_1|^2 \right] 
+ \left[ \sum_{\substack{i=2}}^{N} \alpha_i \ket{b_i} - \alpha_1 \ket{b_1} \right]\\
&=   -\left[ 2 - 4|\alpha_1|^2 \right] \ket{\psi}
+ \sum_{\substack{i=2}}^{N} \alpha_i \ket{b_i} - \alpha_1 \ket{b_1}\\
&= -\left[ 2 - 4|\alpha_1|^2 -1 \right]\sum_{\substack{i=2}}^{N} \alpha_i \ket{b_i} - \left[ 2 - 4|\alpha_1|^2 +1 \right]\alpha_1 \ket{b_1}\\
\end{align*}
\end{widetext}
As we assume $\ket{b_1}$ is the secret, therefore
$$|\braket{b_1}{U_{\ket{\psi}}U_{\ket{s}}|{\psi}}|^2 =|\alpha_1|^2(3-4|\alpha_1|^2)^2 $$

For the forward direction, suppose that the scheme is perfectly recoverable. Then,
$$|\braket{b_1}{U_{\ket{\psi}}U_{\ket{s}}|{\psi}}|^2 = 1.$$
$$\therefore |\alpha_1|^2(3-4|\alpha_1|^2)^2=1 \equiv |\alpha_1|(3-4|\alpha_1|^2) = \pm 1$$

Hence, the possible values for \( |\alpha_1| \) are \( \frac{1}{2} \) and \( 1 \). However, we discard \( |\alpha_1| = 1 \) since it results in the state \( \ket{b_1} \) only. The only possibility is 
$$|\bra{b_1}\ket{\psi}|^2 = \tfrac{1}{4}, \quad \mathrm{or,}\quad |\bra{b_1}\ket{\psi}|=\tfrac{1}{2}, $$
as required.

Conversely, let $|\bra{b_1}\ket{\psi}|^2 = |\alpha_1|^2 =\frac{1}{4} $, where $\ket{b_1}$ is the secret state.
Now, 
$$|\braket{b_1}{U_{\ket{\psi}}U_{\ket{s}}|{\psi}}|^2 =|\alpha_1|^2(3-4|\alpha_1|^2)^2 $$
Putting $|\alpha_1|^2 =\frac{1}{4} $ gives us 
$$|\braket{b_1}{U_{\ket{\psi}}U_{\ket{s}}|{\psi}}|^2 =1, $$
implying that the scheme is perfectly recoverable.
\end{proof}

\section{Appendix: Hsu's Nonces~\label{appendix:hsu_sigma_table}} 
Hsu introduced the use of nonces~\cite{hsu-scheme} by selecting $\ket{A} \otimes \ket{B} \in \{\ket{+}, \ket{-}, \ket{+i}, \ket{-i}\}^{\otimes 2}$.
Below, we list the states \(U_s\ket{\psi_i}\) for all \(i\).
For clarity, we classify the possible choices into four cases:\\
(i) \(\ket{A} \in \{\ket{+}, \ket{-}\}\) and \(\ket{B} \in \{\ket{+}, \ket{-}\}\); \\
(ii) \(\ket{A} \in \{\ket{+}, \ket{-}\}\) and \(\ket{B} \in \{\ket{+i}, \ket{-i}\}\); \\
(iii) \(\ket{A} \in \{\ket{+i}, \ket{-i}\}\) and \(\ket{B} \in \{\ket{+}, \ket{-}\}\); and \\
 (iv) \(\ket{A} \in \{\ket{+i}, \ket{-i}\}\) and
\(\ket{B} \in \{\ket{+i}, \ket{-i}\}\).



\begin{table}[h]
\centering
\renewcommand{\arraystretch}{1.2}
\small
\begin{tabular}{|c | c | c | c | c|}
\toprule
\textbf{secret} & $\mathbf{\ket{+}\ket{+}}$ & $\ket{+}\ket{-}$ & $\ket{-}\ket{+}$ & $\ket{-}\ket{-}$\\
\midrule
$s=00$ & $\frac{1}{\sqrt{2}} (-\ket{0})\ket{-} + \frac{1}{\sqrt{2}}\ket{1})\ket{+}$  & $\frac{1}{\sqrt{2}} (-\ket{0}) \ket{+} + \frac{1}{\sqrt{2}} \ket{1}\ket{-}$ & $\frac{1}{\sqrt{2}} (-\ket{0}) \ket{-} + \frac{1}{\sqrt{2}} (-\ket{1})\ket{+}$ & $\frac{1}{\sqrt{2}} (-\ket{0}) \ket{+} + \frac{1}{\sqrt{2}} (-\ket{1})\ket{-}$\\

$s=01$ &  $\frac{1}{\sqrt{2}} (\ket{0})\ket{-} + \frac{1}{\sqrt{2}}\ket{1}\ket{+}$ & $\frac{1}{\sqrt{2}} \ket{0} \ket{+} + \frac{1}{\sqrt{2}} \ket{1}\ket{-}$ & $\frac{1}{\sqrt{2}} \ket{0} \ket{-} + \frac{1}{\sqrt{2}} (-\ket{1})\ket{+}$ & $\frac{1}{\sqrt{2}} \ket{0} \ket{+} + \frac{1}{\sqrt{2}} (-\ket{1})\ket{-}$\\

$s=10$ & $\frac{1}{\sqrt{2}} \ket{0} \ket{+} + \frac{1}{\sqrt{2}} (-\ket{1})\ket{-}$ & $\frac{1}{\sqrt{2}} \ket{0}\ket{-} + \frac{1}{\sqrt{2}}(-\ket{1})\ket{+}$ & $\frac{1}{\sqrt{2}} \ket{0} \ket{+} + \frac{1}{\sqrt{2}} \ket{1}\ket{-}$& $\frac{1}{\sqrt{2}} \ket{0} \ket{-} + \frac{1}{\sqrt{2}} \ket{1}\ket{+}$\\

$s=11$ & $\frac{1}{\sqrt{2}} \ket{0} \ket{+} + \frac{1}{\sqrt{2}} \ket{1} \ket{-}$ &  $\frac{1}{\sqrt{2}} \ket{0}\ket{-} + \frac{1}{\sqrt{2}}\ket{1}\ket{+}$ & $\frac{1}{\sqrt{2}} \ket{0} \ket{+} + \frac{1}{\sqrt{2}} (-\ket{1})\ket{-}$ & $\frac{1}{\sqrt{2}} \ket{0} \ket{-} + \frac{1}{\sqrt{2}} (-\ket{1})\ket{+}$\\
\bottomrule
\end{tabular}
\end{table}

\begin{table}[h!]
\centering
\renewcommand{\arraystretch}{1.2}
\small
\begin{tabular}{|c | c | c | c | c|}
\toprule
\textbf{secret} & ${\ket{+}\ket{+i}}$ & $\ket{+}\ket{-i}$ & $\ket{-}\ket{+i}$ & $\ket{-}\ket{-i}$\\
\midrule
$s=00$ & $\frac{1}{\sqrt{2}} (-\ket{0})\ket{-i} + \frac{1}{\sqrt{2}}\ket{1})\ket{+i}$  & $\frac{1}{\sqrt{2}} (-\ket{0}) \ket{+i} + \frac{1}{\sqrt{2}} \ket{1}\ket{-i}$ & $\frac{1}{\sqrt{2}} (-\ket{0}) \ket{-i} + \frac{1}{\sqrt{2}} (-\ket{1})\ket{+i}$ & $\frac{1}{\sqrt{2}} (-\ket{0}) \ket{+} + \frac{1}{\sqrt{2}} (-\ket{1})\ket{-}$\\

  $s=01$ &  $\frac{1}{\sqrt{2}} (\ket{0})\ket{-i} + \frac{1}{\sqrt{2}}\ket{1}\ket{+i}$ & $\frac{1}{\sqrt{2}} \ket{0} \ket{+i} + \frac{1}{\sqrt{2}} \ket{1}\ket{-i}$ & $\frac{1}{\sqrt{2}} \ket{0} \ket{-i} + \frac{1}{\sqrt{2}} (-\ket{1})\ket{+i}$ & $\frac{1}{\sqrt{2}} \ket{0} \ket{+i} + \frac{1}{\sqrt{2}} (-\ket{1})\ket{-i}$\\

  $s=10$ & $\frac{1}{\sqrt{2}} \ket{0} \ket{+i} + \frac{1}{\sqrt{2}} (-\ket{1})\ket{-i}$ & $\frac{1}{\sqrt{2}} \ket{0}\ket{-i} + \frac{1}{\sqrt{2}}(-\ket{1})\ket{+i}$ & $\frac{1}{\sqrt{2}} \ket{0} \ket{+i} + \frac{1}{\sqrt{2}} \ket{1}\ket{-i}$& $\frac{1}{\sqrt{2}} \ket{0} \ket{-i} + \frac{1}{\sqrt{2}} \ket{1}\ket{+i}$\\

$s=11$ & $\frac{1}{\sqrt{2}} \ket{0} \ket{+} + \frac{1}{\sqrt{2}} \ket{1} \ket{-i}$ &  $\frac{1}{\sqrt{2}} \ket{0}\ket{-i} + \frac{1}{\sqrt{2}}\ket{1}\ket{+i}$ & $\frac{1}{\sqrt{2}} \ket{0} \ket{+i} + \frac{1}{\sqrt{2}} (-\ket{1})\ket{-i}$ & $\frac{1}{\sqrt{2}} \ket{0} \ket{-i} + \frac{1}{\sqrt{2}} (-\ket{1})\ket{+i}$\\
\bottomrule
\end{tabular}
\end{table}

\begin{table}[h!]
\centering
\renewcommand{\arraystretch}{1.2}
\small
\begin{tabular}{|c | c | c | c | c|}
\toprule
\textbf{secret} & ${\ket{+i}\ket{+}}$ & $\ket{+i}\ket{-}$ & $\ket{-i}\ket{+}$ & $\ket{-i}\ket{-}$\\
\midrule
$s=00$ & $\frac{1}{\sqrt{2}} (-\ket{0})\ket{-} + \frac{1}{\sqrt{2}}(i\,\ket{1})\ket{+}$  & $\frac{1}{\sqrt{2}} (-\ket{0}) \ket{+} + \frac{1}{\sqrt{2}} (i\,\ket{1})\ket{-}$ & $\frac{1}{\sqrt{2}} (-\ket{0}) \ket{-} + \frac{1}{\sqrt{2}} (-i\ket{1})\ket{+}$ & $\frac{1}{\sqrt{2}} (-\ket{0}) \ket{+} + \frac{1}{\sqrt{2}} (-i\ket{1})\ket{-}$\\

$s=01$ &  $\frac{1}{\sqrt{2}} (\ket{0})\ket{-} + \frac{1}{\sqrt{2}}(i\,\ket{1})\ket{+}$ & $\frac{1}{\sqrt{2}} \ket{0} \ket{+} + \frac{1}{\sqrt{2}} (i\,\ket{1})\ket{-}$ & $\frac{1}{\sqrt{2}} \ket{0} \ket{-} + \frac{1}{\sqrt{2}} (-i\ket{1})\ket{+}$ & $\frac{1}{\sqrt{2}} \ket{0} \ket{+} + \frac{1}{\sqrt{2}} (-i\ket{1})\ket{-}$\\

$s=10$ & $\frac{1}{\sqrt{2}} \ket{0} \ket{+} + \frac{1}{\sqrt{2}} (-i\,\ket{1})\ket{-}$ & $\frac{1}{\sqrt{2}} \ket{0}\ket{-} + \frac{1}{\sqrt{2}}(-i\,\ket{1})\ket{+}$ & $\frac{1}{\sqrt{2}} \ket{0} \ket{+} + \frac{1}{\sqrt{2}} (i\,\ket{1})\ket{-}$& $\frac{1}{\sqrt{2}} \ket{0} \ket{-} + \frac{1}{\sqrt{2}} (i\ket{1})\ket{+}$\\

$s=11$ & $\frac{1}{\sqrt{2}} \ket{0} \ket{+} + \frac{1}{\sqrt{2}} (i\,\ket{1}) \ket{-}$ &  $\frac{1}{\sqrt{2}} \ket{0}\ket{-} + \frac{1}{\sqrt{2}}(i\,\ket{1})\ket{+}$ & $\frac{1}{\sqrt{2}} \ket{0} \ket{+} + \frac{1}{\sqrt{2}} (-i\ket{1})\ket{-}$ & $\frac{1}{\sqrt{2}} \ket{0} \ket{-} + \frac{1}{\sqrt{2}} (-i\ket{1})\ket{+}$\\
\bottomrule
\end{tabular}
\end{table}

\begin{table}[h]
\centering
\renewcommand{\arraystretch}{1.2}
\small
\begin{tabular}{|c | c | c | c | c|}
\toprule
\textbf{secret} & ${\ket{+i}\ket{+i}}$ & $\ket{+i}\ket{-i}$ & $\ket{-i}\ket{+i}$ & $\ket{-i}\ket{-i}$\\
\midrule
$s=00$ & $\frac{1}{\sqrt{2}} (-\ket{0})\ket{-i} + \frac{1}{\sqrt{2}}(i\,\ket{1})\ket{+i}$  & $\frac{1}{\sqrt{2}} (-\ket{0}) \ket{+i} + \frac{1}{\sqrt{2}} (i\,\ket{1})\ket{-i}$ & $\frac{1}{\sqrt{2}} (-\ket{0}) \ket{-i} + \frac{1}{\sqrt{2}} (-i\ket{1})\ket{+i}$ & $\frac{1}{\sqrt{2}} (-\ket{0}) \ket{+i} + \frac{1}{\sqrt{2}} (-i\ket{1})\ket{-i}$\\

$s=01$ &  $\frac{1}{\sqrt{2}} (\ket{0})\ket{-i} + \frac{1}{\sqrt{2}}(i\,\ket{1})\ket{+i}$ & $\frac{1}{\sqrt{2}} \ket{0} \ket{+i} + \frac{1}{\sqrt{2}} (i\,\ket{1})\ket{-i}$ & $\frac{1}{\sqrt{2}} \ket{0} \ket{-i} + \frac{1}{\sqrt{2}} (-i\ket{1})\ket{+i}$ & $\frac{1}{\sqrt{2}} \ket{0} \ket{+i} + \frac{1}{\sqrt{2}} (-i\ket{1})\ket{-i}$\\

$s=10$ & $\frac{1}{\sqrt{2}} \ket{0} \ket{+i} + \frac{1}{\sqrt{2}} (-i\,\ket{1})\ket{-i}$ & $\frac{1}{\sqrt{2}} \ket{0}\ket{-i} + \frac{1}{\sqrt{2}}(-i\,\ket{1})\ket{+i}$ & $\frac{1}{\sqrt{2}} \ket{0} \ket{+i} + \frac{1}{\sqrt{2}} (i\,\ket{1})\ket{-i}$& $\frac{1}{\sqrt{2}} \ket{0} \ket{-i} + \frac{1}{\sqrt{2}} (i\ket{1})\ket{+i}$\\

$s=11$ & $\frac{1}{\sqrt{2}} \ket{0} \ket{+i} + \frac{1}{\sqrt{2}} (i\,\ket{1}) \ket{-i}$ &  $\frac{1}{\sqrt{2}} \ket{0}\ket{-i} + \frac{1}{\sqrt{2}}(i\,\ket{1})\ket{+i}$ & $\frac{1}{\sqrt{2}} \ket{0} \ket{+i} + \frac{1}{\sqrt{2}} (-i\ket{1})\ket{-i}$ & $\frac{1}{\sqrt{2}} \ket{0} \ket{-i} + \frac{1}{\sqrt{2}} (-i\ket{1})\ket{+i}$\\
\bottomrule
\end{tabular}
\end{table}


\end{document}